\documentclass{aastex}
\usepackage{emulateapj5,apjfonts}

\journalinfo{The Astrophysical Journal}

\slugcomment{Received 2003 Aug 4; Accepted 2003 Nov 7}
 
\newcommand\kms{\ifmmode {\rm km\ s}^{-1} \else km s$^{-1}$\fi} 
\newcommand\ctssec{\ifmmode {\rm cts\ s}^{-1} \else cts s$^{-1}$\fi}  
\newcommand\ergsec{\ifmmode {\rm ergs\ s}^{-1} \else  ergs s$^{-1}$\fi}  
\newcommand\eflux{\ifmmode {\rm ergs\ s}^{-1}\;{\rm cm}^{-2} 
	\else  ergs s$^{-1}$ cm$^{-2}$\fi}  
\newcommand\phflux{\ifmmode {\rm photons\ s}^{-1}\;{\rm cm}^{-2} 
	\else  photons s$^{-1}$ cm$^{-2}$\fi}  
\newcommand\esfluxhz{\ifmmode {\rm ergs\ s}^{-1}\;{\rm cm}^{-2} Hz$^{-1}$
	\else  ergs s$^{-1}$ cm$^{-2}$ Hz$^{-1}$\fi}
\newcommand\esfluxa{\ifmmode {\rm ergs\ s}^{-1}\;{\rm cm}^{-2} \AA$^{-1}$
	\else  ergs s$^{-1}$ cm$^{-2}$ \AA$^{-1}$\fi}
\newcommand\cc{\ifmmode {\rm cm}^{-3} \else cm$^{-3}$\fi}  
\newcommand\FWHM{\ifmmode {\rm FWHM} \else ${\rm FWHM}$\fi}  
\newcommand\Msun{\ifmmode M_{\odot} \else $M_{\odot}$\fi} 
\newcommand\Lsun{\ifmmode L_{\odot} \else $L_{\odot}$\fi} 
\newcommand\ltsim{\raisebox{-.5ex}{$\;\stackrel{<}{\sim}\;$}} 
\newcommand\gtsim{\raisebox{-.5ex}{$\;\stackrel{>}{\sim}\;$}} 
\newcommand\Hbeta{\ifmmode {\rm H}\beta \else H$\beta$\fi} 
\newcommand\Kalpha{\ifmmode {\rm K}\alpha \else K$\alpha$\fi} 
\newcommand\NH{\ifmmode N_{\rm H} \else N$_{\rm H}$\fi} 
 
\shorttitle{SED of NLS1 Arakelian 564} 
\shortauthors{Romano et al.\ 2002} 
\received{2003 Aug 04} 
\accepted{2003 Nov 07} 
\begin{document} 
\title{The Spectral Energy Distribution and Emission-Line 
	properties \\ 
	 of the NLS1 Galaxy Arakelian~564}
\author{P.\ Romano\altaffilmark{1,2}, 
	S. Mathur\altaffilmark{1},
	T.J. Turner\altaffilmark{3,4},
	S.B. Kraemer\altaffilmark{5},
	D.M. Crenshaw\altaffilmark{6}, 
 	B.M. Peterson\altaffilmark{1},
	R.W. Pogge\altaffilmark{1}, 
	W.N. Brandt\altaffilmark{7},
	I.M. George\altaffilmark{3,4},
	K. Horne\altaffilmark{8},
	G.A. Kriss\altaffilmark{9},
	H. Netzer\altaffilmark{10},
	O. Shemmer\altaffilmark{10},
	and W. Wamsteker\altaffilmark{11} 
}
\altaffiltext{1}{Department of Astronomy, The Ohio State University,  
	140 West 18th Avenue, Columbus, OH  43210.
		}
\altaffiltext{2}{Current address: INAF--Osservatorio Astronomico di Brera, 
	Via E.\ Bianchi 46, 23807 Merate (LC), Italy; romano@merate.mi.astro.it.}
\altaffiltext{3}{Laboratory for High Energy Astrophysics, Code 660, NASA's 
	Goddard Space Flight Center, Greenbelt, MD 20771.}
\altaffiltext{4}{Joint Center for Astrophysics, Physics Department, University 
	of Maryland, Baltimore County, 1000 Hilltop Circle, Baltimore, MD 21250.}
\altaffiltext{5}{Catholic University of America and Laboratory for Astronomy
	and Solar Physics, NASA's Goddard Space Flight Center, Code 681,
	Greenbelt, MD  20771.}
\altaffiltext{6}{Department of Physics and Astronomy, Georgia State 
	University, Astronomy Offices, One Park Place South SE, Suite 700,
	Atlanta, GA 30303.}
\altaffiltext{7}{Department of Astronomy \& Astrophysics, The Pennsylvania 
	State University, 525 Davey Laboratory, University Park, PA 16802.}
\altaffiltext{8}{School of Physics and Astronomy, University of St. Andrews, 
	St. Andrews, KY16 9SS, UK.}
\altaffiltext{9}{Space Telescope Science Institute, 3700 San Martin Drive, 
	Baltimore, MD 21218.}
\altaffiltext{10}{School of Physics and Astronomy and the Wise Observatory, 
	The Raymond and Beverly Sackler Faculty of Exact Sciences, 
	Tel Aviv University, Tel Aviv 69978, Israel.}
\altaffiltext{11}{ESA, P.O. Box 50727, 28080 Madrid, Spain.}

	\begin{abstract} 
We present the intrinsic spectral energy distribution (SED) of the 
Narrow-Line Seyfert 1 galaxy (NLS1) \objectname[MGC +05-53-012]{Ark~564}, 
constructed with contemporaneous data obtained during a 
multi-wavelength, multi-satellite observing campaign in 
2000 and 2001. 
We compare this SED with that of the NLS1 Ton~S180 
and with those obtained for Broad-Line Seyfert 1s to infer how the 
relative accretion rates vary among the Seyfert 1 population. 
Although the peak of the SED is not well constrained, 
in our parameterization most of the energy of this object is 
emitted in the 10--100\,eV regime, constituting roughly half 
of the emitted energy in the optical/X-ray ranges. 
This is consistent with a primary spectral component peaking 
in the extreme UV/soft X-ray band, and with disk--corona models, 
hence high accretion rates. 
Indeed, we  estimate that  $\dot{m} \approx 1$.
We also address the issue of the energy budget in this source
by examining the emission lines observed in its spectrum, 
and we constrain the physical properties of the line-emitting gas 
through photoionization modeling. 
The available data suggest that the line-emitting gas is characterized by 
$\log n \approx 11$ and $\log U \approx 0$, and is stratified
around $\log U \approx 0$.
Our estimate of the radius of the H$\beta$-emitting region 
$R_{\rm BLR}^{{\rm H}\beta} \approx 10 \pm 2$ lt-days  
is consistent with the $R_{\rm BLR}^{{\rm H}\beta}$--luminosity 
relationships found for Sy1 galaxies, which indicates that the 
narrowness of the emission lines in this NLS1 
is not due to the Broad-Line Region  being relatively 
further away from the central mass than in 
BLS1s of comparable luminosity. 
We also find evidence for super-solar metallicity in this NLS1. 
We show that the emission lines are not good diagnostics for
the underlying SEDs and that the absorption line studies 
offer a far more powerful tool to determine the 
ionizing continuum of AGNs, especially if comparing 
the lower- and higher-ionization lines.
	\end{abstract}


	\section{Introduction}	

The population of Seyfert 1 galaxies has a widely used 
sub-classification into Narrow-Line Seyfert 1 galaxies (NLS1s) 
and Broad-Line Seyfert 1 galaxies (BLS1s). While this
classification appears to make an arbitrary distinction based on  
the widths of the optical emission lines (NLS1s having FWHM(H$\beta) 
\lesssim 2000$ ${\rm km\ s^{-1}}$, \citealt{Goodrich89}), 
this is in fact an extremely useful scheme since 
the X-ray properties of the two subclasses are systematically different.
As a class, NLS1s show rapid and large-amplitude X-ray variability 
\citep[][hereafter BBF96; \citealt{Turnerea99b}]{BBF96}, with the excess variance 
\citep{Nandraea97a} typically an order of magnitude larger than that 
observed for samples of BLS1s  with the same luminosity distribution 
\citep{Turnerea99b,Leighly99I}. Analogously, the spectral properties 
also vary across the Seyfert population with 
NLS1s showing systematically steeper spectra 
than those of BLS1s in both the soft and hard X-ray bands 
\citep{BBF96,BME97,TGN98,Leighly99II,Vaughan99b}. 

One increasingly popular hypothesis to explain the 
differences in X-ray properties across the Seyfert population is that 
NLS1s have relatively low masses for the central black hole
compared to BLS1s with similar luminosities.
Smaller black-hole masses naturally explain both the narrowness of
the optical emission lines, which are generated in gas that has
relatively small Keplerian velocities, and the extreme
X-ray variability, since the primary emission would originate in a
smaller region around the central engine \citep[e.g.,][]{Laorea97}.
Given that NLS1s have comparable
luminosity to that of the BLS1s, \citet{PDO95} suggested that 
they must be emitting at higher fractions of their Eddington luminosity, 
hence higher fractional accretion rates 
($\dot{m} = \dot{M}/\dot{M}_{\mbox{\scriptsize Edd}}$) 
are also required. 
The closer the luminosity is to the
Eddington limit (and the lower the black-hole mass), the greater the
fraction of the energy emitted by the accretion disk in the soft
X-rays \citep{RFM92}. Thus NLS1s might be expected to show 
disk components which peak at higher energies than for BLS1s. 
\citet{PDO95} and \citet{MH97} noted that soft photons from the disk may 
Compton-cool hard X-rays from the corona, and 
cause the observed steep spectra. 

Given the dependence of the peak energy of the disk emission on the 
accretion rate, the spectral energy distribution (SED) of an active galactic 
nucleus (AGN) will provide critical information about accretion rates and 
conditions close to the disk. In particular, examination of the SED of a 
NLS1, and comparison with that obtained for BLS1s will offer insight into the 
relative accretion rates across the Seyfert population. 
However, measuring the SED requires observations taken simultaneously over a long 
wavelength baseline stretching from infrared to hard X-ray wavelengths,
and the determination of the UV/X-ray continuum in AGN is still difficult
due to the severe attenuation by even small amounts of Galactic interstellar 
gas along the line-of-sight. 
Another complication comes from intrinsic reddening, i.e., reddening 
associated with the active nucleus itself. 
Indeed, the steeper UV/blue continua observed in NLS1s 
(when compared to AGN spectrum composites)
can be attributed at least in part to reddening, though the ionization of the 
absorbing material and its location with respect to the accretion source 
is still not well determined \citep{Constantinea03}.

Arakelian~564 (Ark~564, IRAS 22403+2927, MGC +05-53-012) is a bright, nearby
NLS1 galaxy, with $z = 0.02467$ and  $V = 14.6$ mag
\citep{rc3.9catalogue}, and a mean 2--10\,keV luminosity 
$L_{\mbox{\scriptsize 2--10{} keV}} \approx 2.4 \times 10^{43}$
\ergsec{} with flux variations of a factor of a few in a few thousand seconds 
\citep[][hereafter Paper~I]{Akn564I}. 
It was the object of an intense 
multiwavelength monitoring campaign that included simultaneous 
observations from {\it ASCA} (2000 June 1 to July 6, Paper~I;  
\citealt{Poundsea01,Edelsonea02}), 
{\it XMM-Newton} (2000 June 17, \citealt{Vignaliea03}), 
{\it Chandra} (2000 June 17, \citealt{Matsumotoea02}), 
{\it HST} (2000 May 9 to July 8, \citealt{Collierea01}, Paper~II; 
\citealt{Crenshawea02}, Paper~IV), 
{\it FUSE} (2001 June 29--30, \citealt{Romanoea02b}, Paper V), 
and from many ground-based observatories as part of an  
AGN Watch\footnote{\anchor{http://www.astronomy.ohio-state.edu/~agnwatch}
{All publicly available data and complete references to published AGN Watch 
papers can be found at http://www.astronomy.ohio-state.edu/$\sim$agnwatch.}} 
project \citep[1998 November to 2001 January, ][ Paper~III]{Shemmerea01}. 
Ark~564 has shown a strong associated UV absorber 
(\citealt{Crenshawea99}, Paper~II; Paper~IV; Paper V). 
There are indications that it also possesses a warm X-ray absorber, 
as seen by the narrow absorption lines of \ion{O}{7} and \ion{O}{8} 
detected in a {\it Chandra} spectrum 
\citep{Matsumotoea02}, and that the UV and X-ray absorbers 
in Arakelian~564 are physically related, possibly identical,
and may be spatially extended along the line of sight (Paper~V).

In this paper we present a contemporaneous SED of Ark~564, based on 
the extensive monitoring of 2000. 
In \S\ref{sa:allobs} we describe the observations and data reduction, 
summarize the main results of the monitoring campaign, and describe
the adopted method for reddening correction.
In \S\ref{sa:sed} we present the SED of Ark~564.
In \S\ref{sa:pioniza} we constrain the mean physical properties of 
the line-emitting gas through photoionization modeling. 
In \S\ref{sa:discuss} we discuss some implications of our 
investigation. 
Finally, our results are summarized in \S\ref{sa:summary}.

	\section{Observations\label{sa:allobs}} 

	\subsection{Data Reduction\label{sa:dataredux}} 

Table~\ref{sa:obslogs} summarizes the log of the observations of Ark~564
used in this study. 
Column 1 and 2 report the observatory and instrument which obtained the data;
Column 3 shows the date in which the data were obtained; 
Column 4 lists the ranges of energy and wavelength we used for our work;
Column 5 lists the total exposure times and the slit size, along with 
other relevant notes for a particular observation; 
finally, Column 6 reports the references to the papers where the data were 
first published. 
Most of the data have been published already, with the exception of the 
{\it XMM} data which will be published in \citet{Vignaliea03}, 
therefore here we will only 
briefly summarize their reduction and analysis, referring the reader
to the original papers for further details, and to \S\ref{sa:mainres}
for a summary of the main results.

The {\it ASCA} data (Paper~I) were obtained starting from 2000 Jun 1 
(Julian Date 2451697.024) for a total exposure time of 
$\sim 2.98$\,Ms, and were reduced using standard techniques 
as described in \citet{Nandraea97a} with the methods and screening
criteria utilized by the {\it Tartarus}\footnote{
\anchor{http://tartarus.gsfc.nasa.gov}{http://tartarus.gsfc.nasa.gov. }} 
database \citep{Turnerea99b}.
Data screening yielded an effective exposure time of $\sim 1.11$ Ms
for the SISs and $\sim 1.29$ Ms for the GISs. 
The degradation of the low energy response of the SIS detectors
was corrected for with the method of \citet{Yaqoob}\footnote{
\anchor{http://lheawww.gsfc.nasa.gov/$\sim$yaqoob/ccd/nhparam.html}
{see http://lheawww.gsfc.nasa.gov/$\sim$yaqoob/ccd/nhparam.html.}}, 
i.e.\ we  parameterized the efficiency loss with a time-dependent 
absorption ($\NH(\mbox{SIS0}) =7.5 \times
10^{20}$ cm$^{-2}$, $\NH(\mbox{SIS1}) = 1.05 \times
10^{21}$ cm$^{-2}$, Paper~I).

{\it FUSE} observed Ark~564 for $63$\,ks starting from 
2001 June 29 (Paper V). 
The observations consisted of 24 exposures performed in photon 
address (time-tag) mode through the $30\arcsec \times 30\arcsec$ 
low-resolution (LWRS) aperture. 
The data were reduced with {\tt CalFUSE} 
(version 2.0.5)\footnote{
\anchor{http://fuse.pha.jhu.edu/analysis/calfuse.html}{See 
http://fuse.pha.jhu.edu/analysis/calfuse.html.}}
as a single continuous exposure. 
As a result of a high-voltage anomaly and data screening, the effective
on-source times were 41\,ks in Detector 1A, 39\,ks in Detector 1B,
58\,ks in Detector 2A, and 62\,ks in Detector 2B. 
We discarded  from further analysis the SiC1A and SiC1B spectra because of
a high-voltage anomaly, and the LiF1B spectrum since it showed 
wavelength-dependent differences in flux of up to 30--50\,\% 
compared to the LiF1A. 
We rebinned the full spectrum in a linear wavelength 
scale using 0.6\AA{} bins (100 pixels, effective resolution of 20\,\kms).
Comparison with the flux level of the {\it HST} spectrum indicates that 
in June 2001, when the {\it FUSE} spectrum was obtained, the source 
was $\sim 1.3$ times brighter than in May--July 2000, when the {\it HST}
spectrum was obtained. 
We therefore scaled the {\it FUSE} fluxes by 0.75 to match the {\it HST} 
data. As noted in Paper V, this is not inconsistent with a combination 
of effects such as flux intercalibration uncertainties 
and, most importantly, source flux variability.

The {\it HST} data (Paper~II and IV) were obtained 
with the Space Telescope Imaging Spectrograph (STIS) 
in 46 visits between 2000 May 9--July 8 (the first five
visits were separated by intervals of 5 days, the 
remaining by 1 day). The spectra were obtained 
through the 52\arcsec $\times$ 0\farcs5 slit and the
low-resolution G140L and G230L gratings which yield a
spectral resolution of $\sim 1.2$\,\AA{} and 3.2\,\AA{}
in the 1150--1730\,\AA{} and 1570--3150\,\AA{} ranges, respectively. 
The data were reduced using the IDL software 

\centerline{}
\vspace{-1cm}

\centerline{\includegraphics[width=10.5cm,height=10.0cm]{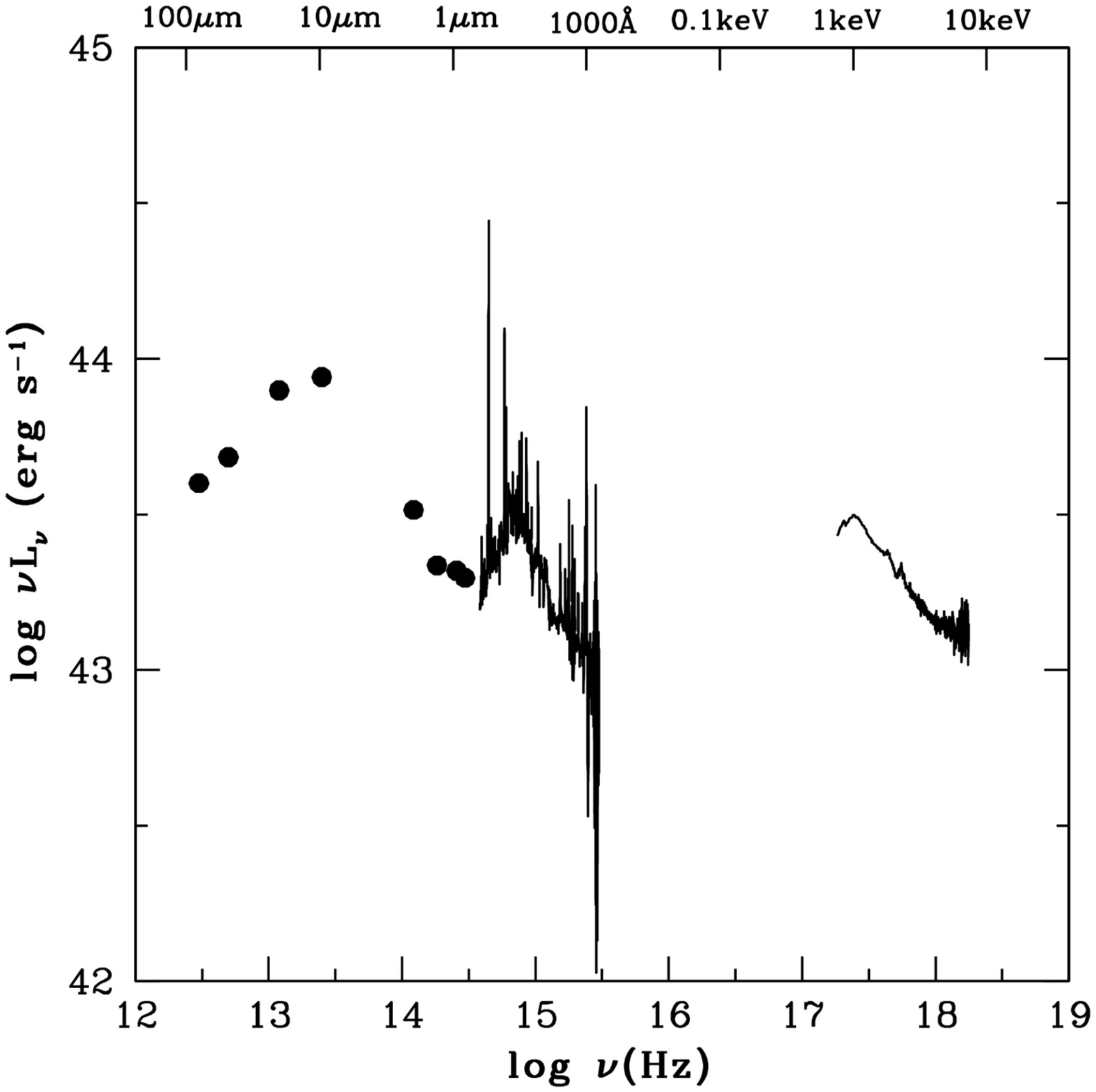}}
\figcaption{Quasi-simultaneous SED of Ark~564 (\S\ref{sa:dataredux}). 
The data have not been corrected for reddening and are in the observed frame.
 \label{sa:obssed}}

\centerline{}
\vspace{0.3cm}

\noindent
developed at NASA's Goddard Space Flight
Center for the STIS Instrument Definition Team \citep{Lindler98}. 
The spectra have been corrected for small 
wavelength intercalibration uncertainties following 
\citet{Koristaea95}.
The uncertainty in the relative wavelength calibration is
on the order of 0.6\,\AA{} and 1.7\,\AA{} for the G140L and 
G230L gratings, respectively. 
A separate mean spectrum was created for the G140L and 
G230L grating separately, given the different resolutions. 

In the optical we combined two spectra.
The first one, which covers the 3170--4160\,\AA{} wavelength range 
with a mean spectral resolution of 0.62\,\AA,  
was obtained in 1980 at Lick Observatory 
(D.\ E.\ Osterbrock 2002, private communication). 
The second one is the mean of the spectra 
taken between 1998 Nov to 2001 Jan at the Tel Aviv 
University Wise Observatory (Paper~III, resolution of $\sim $10\,\AA).
The host galaxy starlight contribution has been estimated by 
measuring its flux through PSF fitting to field stars in
V-band images of the galaxy taken at Wise Observatory, which 
corresponds to $\sim$ 40\,\% of the total light at 
5200\,\AA, i.e.\ $F_{\rm gal} = 2.4 \times 10^{-15}$\,\esfluxa 
(Paper~III). 
Given the limited resolution (dominated by a seeing disk of 
$\sim 2.\arcsec5$), it was not possible to separate the components 
of the host galaxy (bulge and bar) from PSF, hence we 
subtracted from the mean spectrum a constant host contribution 
of $F_{\rm gal}$.
No scaling was necessary between the two spectra, in agreement with the
low-amplitude variations of the continuum found in 
Paper II ($\sim 6$\% over a month-long observation) and 
III ($\sim 10$\% during a two-year monitoring). 
Indeed, there is growing evidence that while NLS1s show very rapid and giant
X-ray variability, they show only slow and minor optical variability
(O.\ Shemmer et al 2003, in preparation).

To extend the SED in the IR, we derived four continuum points 
between $\sim 10000$ and 24000\,\AA{} from 
\citet{rodriguezea02a,rodriguezea02b}, 
which were obtained on 2000 Oct 11 and 13 
with the NASA 
3m {IRTF} telescope and the SPEX spectrome- 

\centerline{}
\vspace{-1cm}

\centerline{\includegraphics[width=10.5cm,height=10.0cm]{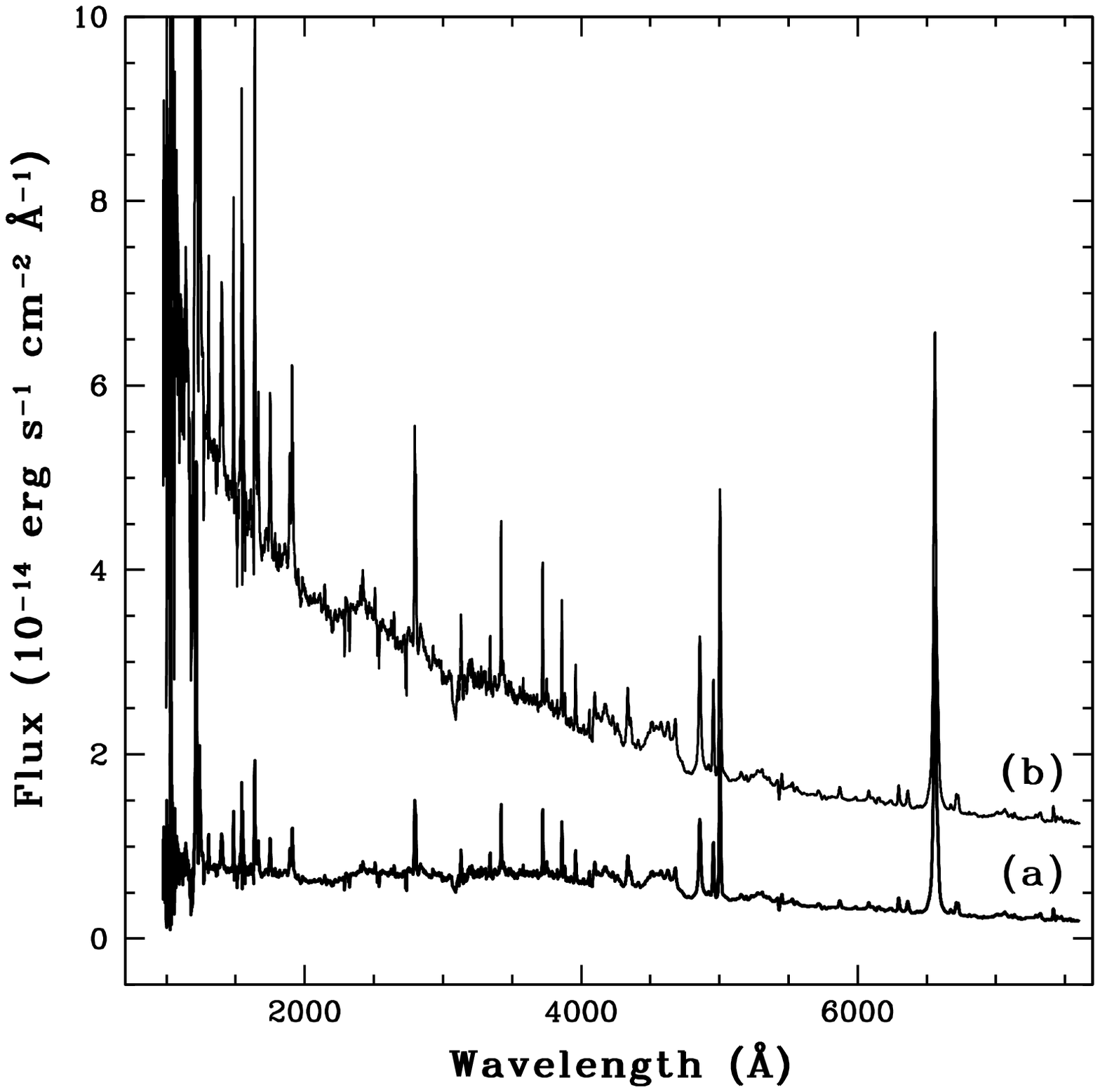}}
\figcaption{{\bf (a)}: 
FUV/Optical rest-frame spectrum of Ark~564, obtained combining 
the {\it FUSE} spectrum (June 2001; 1000-1175\,\AA), 
the {\it HST} G140L and G230L mean spectra (May--July 2000; 1175--3143\,\AA), 
the Lick spectrum (1980; 3170--4160\,\AA), and the Wise mean spectrum 
(1998 Nov--2001 Jan; 4160--7790\,\AA). The {\it FUSE} spectrum has 
been scaled by a factor of 0.75 to match the flux level of the {\it HST} 
spectrum;  the Wise spectrum has been corrected for the galaxy contribution 
($F_{\rm gal} = 2.4 \times 10^{-15}$\,\esfluxa; \citealt{Shemmerea01}). 
{\bf (b)}: Full spectrum after correction for reddening 
using a standard Galactic extinction curve with $E(B-V)=0.03$\,mag plus 
the intrinsic extinction curve that \citet{Crenshawea02} derive 
for Ark~564 and $E(B-V)=0.14$\,mag. 
The data have been corrected for redshift ($z = 0.02467$; 
\citealt{rc3.9catalogue}). The extinction-corrected spectrum 
{\bf (b)} has been offset by $1\times 10^{-14}$\,\esfluxa{}
for clarity. 
\label{sa:optfuv}}

\centerline{}
\vspace{0.3cm}

\noindent
ter. We also retrieved archival {\it IRAS} flux measurements at 
12, 
25, 60, and 100\,$\mu$m \citep{iras} through the 
NASA/IPAC Extragalactic Database (NED). 

Figure~\ref{sa:obssed} shows the contemporaneous SED of Ark~564
before correction for intervening (and intrinsic, see~\S\ref{sa:reddcorr}) 
absorption is applied. 
We note that while the {\it HST} and Wise spectra are simultaneous 
(as well as simultaneous with the {\it ASCA} spectrum), the 
{\it FUSE} and Lick spectra were obtained one year later, and 20 
years earlier, respectively. 
The FUV/optical rest-frame spectrum of Ark~564 
covering the 1000-7790\,\AA{} wavelength region is presented 
in Figure~\ref{sa:optfuv} (labeled as (a)).

	\subsection{Summary of results from the multi-waveband
       observations\label{sa:mainres}} 

The continuum fit to the mean {\it ASCA} spectrum 
(with a power-law model modified by Galactic absorption,   
$N_{\rm H} = 6.4 \times 10^{20}$ cm$^{-2}$, \citealt{DickeyL90}) 
yields a slope $\Gamma = 2.538 \pm 0.005$ (Paper I). 
The strong excess of emission observed below 2\,keV was parameterized
as a Gaussian of peak energy $E=0.57\pm0.02$ keV and 
mean equivalent width (EW)$=110^{+11}_{-15}$\,eV. The soft hump 
component is also found to be variable in 
flux on timescales as short as 1 day and in shape on  
timescales as short as a few days (Paper I).
Parameterization of the soft excess as a 
black-body  yields a temperature $T = 1.8\times 10^{6}$\,K and 
luminosity $L_{\rm bb} = 2.48 \times 10^{38}$\,\ergsec 
(Paper I). 
A strong, ionized ($E \approx 7$\,keV) Fe K$\alpha$ line 
is detected, which shows variations in flux and EW on 
timescales as short as a week (Paper I). %
\centerline{\includegraphics[width=10.5cm,height=10.0cm]{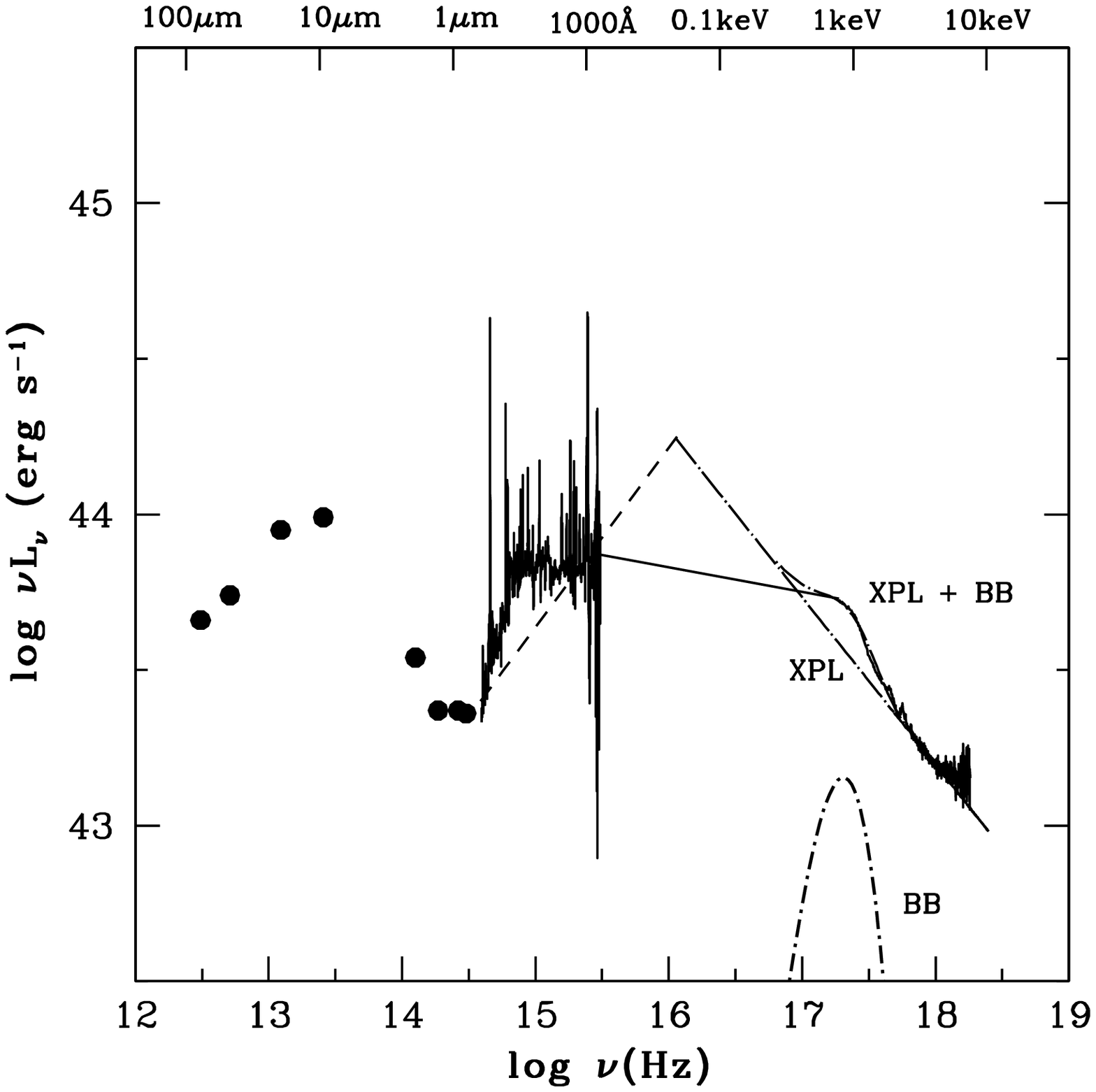}}
\figcaption{Reddening 
corrected, rest-frame SEDs of Ark~564 (see \S\ref{sa:sed}). 
The optical/FUV data are the spectrum labeled (b) in Figure~\ref{sa:optfuv}. 
The X-ray data are drawn from Paper I, while the IR data are from 
{\it IRAS} and {\it IRTF} (\S\ref{sa:dataredux}). 
The short-dashed line shows the power-law fit of the optical/FUV data 
(spectral index $\alpha =0.42\pm0.01$) extrapolated to higher energies. 
The long-dash--dot line is the power-law fit to hard X-ray data 
(XPL, $\alpha_{ASCA} =1.538\pm 0.005$, \S\ref{sa:mainres}) 
extrapolated to lower energies. 
The short-dash--dot line is the black-body model for the soft X-ray excess 
(BB, \S\ref{sa:mainres}). 
The solid line is a more conservative spectral energy distribution 
(SEDB) that connects the FUV and the soft X-ray data with a simple power law 
($\alpha = 1.08$, \S\ref{sa:sed}). 
\label{sa:sedcorr}}

\centerline{}
\vspace{0.3cm}

The {\it FUSE} spectrum is dominated by the strong emission in the 
\ion{O}{6}\,$\lambda\lambda1032, 1038$ resonance doublet, and  
 heavily saturated absorption due to Hydrogen Lyman-Werner bands, 
\ion{O}{6} and \ion{C}{3} are observed at velocities near the 
systemic redshift of Ark~564 (Paper V). 
The available data suggest that the UV and X-ray absorbers 
are physically related, possibly identical, and spatially extended 
along the line of sight, and characterized by total column density 
and ionization parameter 
$\log N_{\rm H}$ (cm$^{-2}) \approx 21$ 
and $\log U \approx -1.5$. 
The absorbing gas is in a state of outflow with respect to the 
nucleus and carries out a kinetic luminosity about one 
order of magnitude smaller than the observed radiative luminosity 
of the source (Paper V). 

Variations in the UV continuum flux (1365\,\AA, Paper~II) are
well correlated with variations in 
both the hard (2-10\,keV) and soft (0.75-2\,keV) X-ray fluxes and with
the soft hump flux. No significant lags are detected between
the variations in the X-ray and UV bands (Paper I). 
The variations of the continuum at 3000\,\AA{} (Paper~II)  
and at 4900\,\AA{} (Paper~III) lag behind those   
at 1365\,\AA{} by $\sim 1$ day and 1.8 days, respectively.
These UV/optical delays were interpreted as evidence for a 
stratified continuum reprocessing region, possibly an accretion 
disk. 

The variations of the Ly$\alpha$ emission line,
which lag the variations of the continuum at 1365\,\AA{} by 3 days,
were used, in conjunction with the line width, to determine the virial 
mass of the central black hole, $M \ltsim 8 \times 10^{6}\,\Msun$
(Paper~II). 
This estimate is uncertain due to the low amplitude of the
Ly$\alpha$ emission line variations (1\%).
However, the estimate in Paper~II 
agrees with the one obtained by \citet{Poundsea01} based on a 
power spectrum analysis of X-ray variability. 
The black hole mass and 5100\,\AA{} luminosity of Ark 564 
are consistent with the hypothesis that NLS1s have lower 
black hole masses and higher accretion rates than BLS1s of comparable  
luminosity. The low level variability observed in the emission 
lines is also different from most Seyfert 1 galaxies, 
which characteristically display variations
of 10\,\% on similar timescales.

	\subsection{Reddening Correction in the Optical/FUV\label{sa:reddcorr}} 

Given the indications (Paper~IV) that strong intrinsic neutral 
absorption is present in Ark~564 in excess of the Galactic 
absorption, special care has been taken in correcting the data 
for reddening. 
We used a standard Galactic extinction curve with $E(B-V)=0.03$\,mag 
plus the intrinsic extinction curve that \citet[][Paper~IV]{Crenshawea02} 
derive for Ark~564 and $E(B-V)=0.14$\,mag. 
The {\it HST} extinction correction was extrapolated linearly into the 
{\it FUSE} band, as suggested by \citet{Hutchingsea01} and 
\citet{Sasseenea02}. 
The effect of reddening correction in the optical/UV bands 
is presented in Figure~\ref{sa:optfuv}, 
where the observed spectrum in the 1000-7790\,\AA{} wavelength region 
(labeled as (a)) is compared to the absorption-corrected one (labeled as (b)).

	\section{The Intrinsic SED of Ark~564\label{sa:sed}}

Figure~\ref{sa:sedcorr} shows the Ark~564 data in the IR/X-ray range.
A power-law fit of the continuum in the optical/FUV 
region\footnote{The fit was made to 9 bands: $\lambda = $ 1005--1007, 
1029.5--1030.5, 1101--1107, 
1114--1118, 1155--1180, 1350--1380, 1460--1500, 1620--1660, and 
7040--7050\,\AA. The uncertainties are purely statistical. 
The continuum fit in Paper~II was performed only 
on the {\it HST} data corrected for Galactic reddening ($E(B-V)=0.06$\, mag) 
and produced $F_{\lambda} \propto \lambda^{-0.88\pm0.01}$;
this is comparable to our value for this reddening, 
$F_{\lambda} \propto \lambda^{-0.73\pm0.01}$, since 
the uncertainties are underestimated in both cases.} 
yields $F_{\lambda} \propto \lambda^{-1.58\pm0.01}$, hence 
spectral index $\alpha =0.42 \pm 0.01$  
(specific flux $f_{\nu} \propto \nu^{-\alpha}$). 
Extrapolation of this power law in the X-ray regime 
greatly overpredicts the X-ray flux (dashed line). 
Analogously, the hard X-ray continuum slope ($\alpha_{ASCA} = \Gamma -1 = 1.538\pm0.005$, 
\S\ref{sa:mainres}, long-dash--dot line) extrapolated to the lower energies 
overpredicts the optical/FUV flux, as previously noted 
by \citet{WalterFink93}.
Clearly, both the optical/FUV and the X-ray power-laws must break 
at some energy between the FUV and soft X-ray. 
With the adopted reddening correction (\S\ref{sa:reddcorr}), the 
spectral energy distribution peaks at $\sim 50$\,eV. 
	
Table~\ref{sa:sedvalues} summarizes some relevant data from the
spectral energy distribution derived using the simple parameterization 
of the combination of an optical/FUV power law with $\alpha =0.42$ breaking 
at $\sim 50$\,eV to $\alpha_{ASCA} = 1.538$ (hereon SEDA), as described above, as well as 
the {\it IRAS} and {\it IRTF} data points (\S\ref{sa:dataredux}).  
Column (1) is the rest wavelength/energy, Column (2) and (3) list the observed 
and reddening corrected values of $\nu\,L_{\nu}$, respectively. 
We estimate that the number of ionizing photons is 
$Q({\rm SEDA}) \approx 10^{55}$ photons~s$^{-1}$.

We also considered a more conservative spectral energy distribution 
(the solid line in Figure~\ref{sa:sedcorr}, hereon SEDB) in which the 
the FUV and the soft X-ray data are connected with a simple power law 
($\alpha = 1.08$), i.e, the combination 
of an optical/FUV power law with $\alpha =0.42$ breaking at 1000\AA{} 
to $\alpha \approx 1$, then again breaking at $\approx 0.8$\,keV to $\alpha_{ASCA} = 1.538$. 
The number of ionizing 
photons for SEDB is $Q({\rm SEDB}) \approx 5 \times 10^{54}$ photons~s$^{-1}$,
which is consistent with what \citet{Crenshawea02} found\footnote{Note 
that SEDB corresponds to SED2 in Paper V.}.

	\section{Photoionization Modeling\label{sa:pioniza}}  

In this section, we address the issue of the energy budget of Ark~564 
by examining the emission lines observed in its spectrum, 
and we deduce the mean physical properties of the line-emitting gas, 
Hydrogen density $n$ 
and ionization parameter $U$ ($U=Q / 4\,\pi \, r^2 \, n\,\,c$, $r$ being the 
distance to the ionizing source and $c$ the speed of light) 
through photoionization modeling given the assumed spectral energy distributions. 
We expect a range of ionization to exist throughout the BLR, to accurately describe which 
multi-zone modeling would be required; 
however, here we use a single density and ionization parameter 
modeling to derive the mean BLR properties.
For each of the input continua (\S\ref{sa:pionizacontinua}), we considered a total 
hydrogen density of $n = 10^{9}, 10^{11}, 10^{12},$ and $10^{13}$ cm$^{-3}$,
and calculated the predicted intensities of the major emission lines, for 
a range of ionization parameters between $\log U = -4$ and $\log U = 2$. 
In the case of ``table agn''(\S\ref{sa:pionizacontinua}), 
we specified a grid of $n$  and $U$ values. 
For the other input continua, we normalized the SEDs  with respect to 
the measured X-ray luminosity in the absorption-corrected rest-frame
2--10\,keV energy range 
($L_{\mbox{\scriptsize 2--10{} keV}} = 2.4 \times 10^{43}$ \ergsec{}; Paper I), 
and specified the radius of the cloud, thus obtaining $U$. 
In the case of SEDA and SEDB, using an observed spectral energy distribution 
assumes that the gas responsible for the emission lines sees 
the same ionizing continuum as the observer does;
therefore, we expect SEDA and SEDB to yield more realistic 
predictions of the emitted line spectrum. 

	\subsection{Emission Line Fluxes\label{sa:fluxmeasur}}

The fluxes (relative to H$\beta$) of the most prominent emission lines 
in the 1150--6817\AA{} wavelength range have been published in Table~2 
of Paper~IV; here we report a selection of them in Table~\ref{sa:elratios}
(Column (2), relative to Ly$\alpha$).
Because of the NLS1 nature of Ark~564, the contribution from the BELR 
and NELR are strongly blended together and those line ratios include both
components. 
The measured fluxes were corrected for reddening using the continuum 
reddening curve \citep{Crenshawea02}, given the similar extinctions 
for the continuum and emission lines; the errors are propagated
in quadrature from the ones listed in Paper~IV, and they include 
photon noise, continuum placement errors, and reddening errors. 

Columns (2) and (3) of Table~\ref{sa:elratios} also report 
the fluxes of \ion{C}{3}$\lambda977$ and the 
\ion{O}{6}$\lambda\lambda$1032,1038 doublet, 
which can help better constrain the value of $n$ and $U$. 
These lines were modeled in Paper~V, hence both a broad and a narrow component
are available (denoted with BEL and NEL, respectively; Columns (2) and (3)). 
In order to compare with the other line ratios, we needed to account 
for the different contribution of BELR and NELR gas to H$\beta$, which we 
estimated as follows. 
We assumed that the ratio of the narrow component of H$\beta$ and 
[\ion{O}{3}]$\lambda5007$ in NGC~5548, i.e., $0.12\pm0.01$ \citep{KCFP98},
can be used for Ark~564; we scaled it to the observed [\ion{O}{3}]$\lambda5007$
flux for Ark~564, F([\ion{O}{3}]$\lambda 5007) = (2.4 \pm 0.1) \times 10^{13}$ \eflux{} 
(Paper~III), obtaining the NEL ratio F(H$\beta) /$F([\ion{O}{3}]$\lambda5007)$ in Ark~564.
Thus, we estimated that roughly 75\% of the total H$\beta$ flux is from the BEL,
and bracketed this value between 50\% and 100\%, given that the  BEL and NEL contributions to 
\ion{C}{3}$\lambda977$  and \ion{O}{6}$\lambda\lambda$1032,1038 are the same, 
and that the total flux is the 
absolute upper limit to the BEL flux.
We notice that \citet{rodriguezea00} found that on average, 50\% of the 
flux of 
the total H$\beta$ is due to emission from the NELR, and that the
F([\ion{O}{3}]$\lambda5007$)/ F(H$\beta$) emitted in the NELR varies from 
1 to 5, which is much lower than our adopted value ($\sim 8.3$). This result is 
sustained by the analysis of \citet{Continiea03}. 
However, \citet{VCVG01} point out that the low 
F([\ion{O}{3}]$\lambda5007$)/ F(H$\beta)$ values found by \citet{rodriguezea00} 
for NLSy1s are due to the fact that they modeled the broad Balmer component 
with a Gaussian rather than a Lorentzian. 
In the analysis of \citet{VCVG01} the F([\ion{O}{3}]$\lambda5007$)/ F(H$\beta$) 
ratios span the range measured in BLSys, which makes our use of a value derived 
from a well-studied BLS1 reasonable. Furthermore, \citet{Nagaoea01} also show that 
F([\ion{O}{1}]$\lambda6300$)/F([\ion{O}{3}]$\lambda5007$),  
F([\ion{O}{3}]$\lambda4363$)/F([\ion{O}{3}]$\lambda5007$) 
are indistinguishable in NLS1s and BLS1s. 
Given this ambiguity, in the following analysis we will note where our 
assumptions for the deconvolution of \ion{O}{6}$\lambda\lambda$1032,1038 and 
\ion{C}{3}$\lambda977$ affect the results. 

As a comparison, Column (4) reports the corresponding values for a mean QSO 
spectrum, which we derived from \citet{Baldwinea95}  by applying the 
reddening correction appropriate for Ark~564; 
the H$\alpha/$Ly$\alpha$ ratio is derived from \citealt{OP85}). 
Column (5) lists the BEL fluxes of the Sy1.5 NGC~5548, 
corrected for NEL contribution and Galactic reddening
(\citealt{KoristaGoad00} and references therein). 
Column (6) lists the FWHM of the lines, as drawn from Papers V, II and III 
(\ion{C}{3}$\lambda 977$ and \ion{O}{6} are the model BEL and NEL components, 
while the others are measured on the whole line profile).
All errors are propagated in quadrature.
Finally, Column (7) reports the references for Columns (2), (3), and (6).
Table~\ref{sa:elratios} shows that Carbon in Ark~564 is at the 
lower end and Nitrogen at the upper end of the mean QSO distribution, 
and some interesting differences can be found with respect to the Sy1.5 NGC~5548. 
Indeed, \ion{N}{5} $\lambda$1240 is stronger in Ark~564 by a 
factor of $\sim 2.3$, while 
\ion{C}{4} $\lambda$1550, \ion{C}{3}] $\lambda$1909, and Mg II $\lambda$2800
are weaker in Ark~564 by a factor of $\sim 4.4$, 2.5 and 2.8, respectively. 

	\subsection{Input Continua\label{sa:pionizacontinua}} 

We used the code 
{\tt Cloudy}\footnote{http://www.nublado.org/.}
\citep[v94.00,][]{Cloudy} to predict the intensities of the lines
produced by the BEL gas through photoionization modeling. 
Our choices of input continua for {\tt Cloudy} are shown in 
Figure~\ref{sa:cloudyseds}.  
In brief,
	\begin{enumerate}
	\item The {\tt Cloudy} ``table agn'' continuum, which is the  \citet{MathewsF87}
continuum modified with a sub-millimeter break at 10\,$\mu$m, so that the 
spectral index is changed from $-1$ to $-5/2$ for frequencies below the 
millimeter break.
	\item The SEDA input continuum, which  was created from points chosen 
from the spectral energy distribution presented in \S\ref{sa:sedcorr} 
(the circles in Figure~\ref{sa:cloudyseds}),
as well as one extrapolated point (the empty square), i.e., 
where the optical/FUV and X-ray power law extrapolations meet. 
In particular, in the X-ray, we used continuum points from the 
power-law fit and added a black body component (the dot-dash line 
in Figure~\ref{sa:sedcorr}) of temperature 
$T = 1.8\times 10^{6}$\,K 
and luminosity $L_{\rm bb} = 2.48 \times 10^{38}$\,\ergsec 
(\S\ref{sa:mainres}). 
	\item The more conservative SEDB input continuum, which only uses 
points from the observed spectral energy distribution (circles only).
\end{enumerate}

\centerline{}
\vspace{-1.2cm}

\centerline{\includegraphics[width=10.5cm,height=9.5cm]{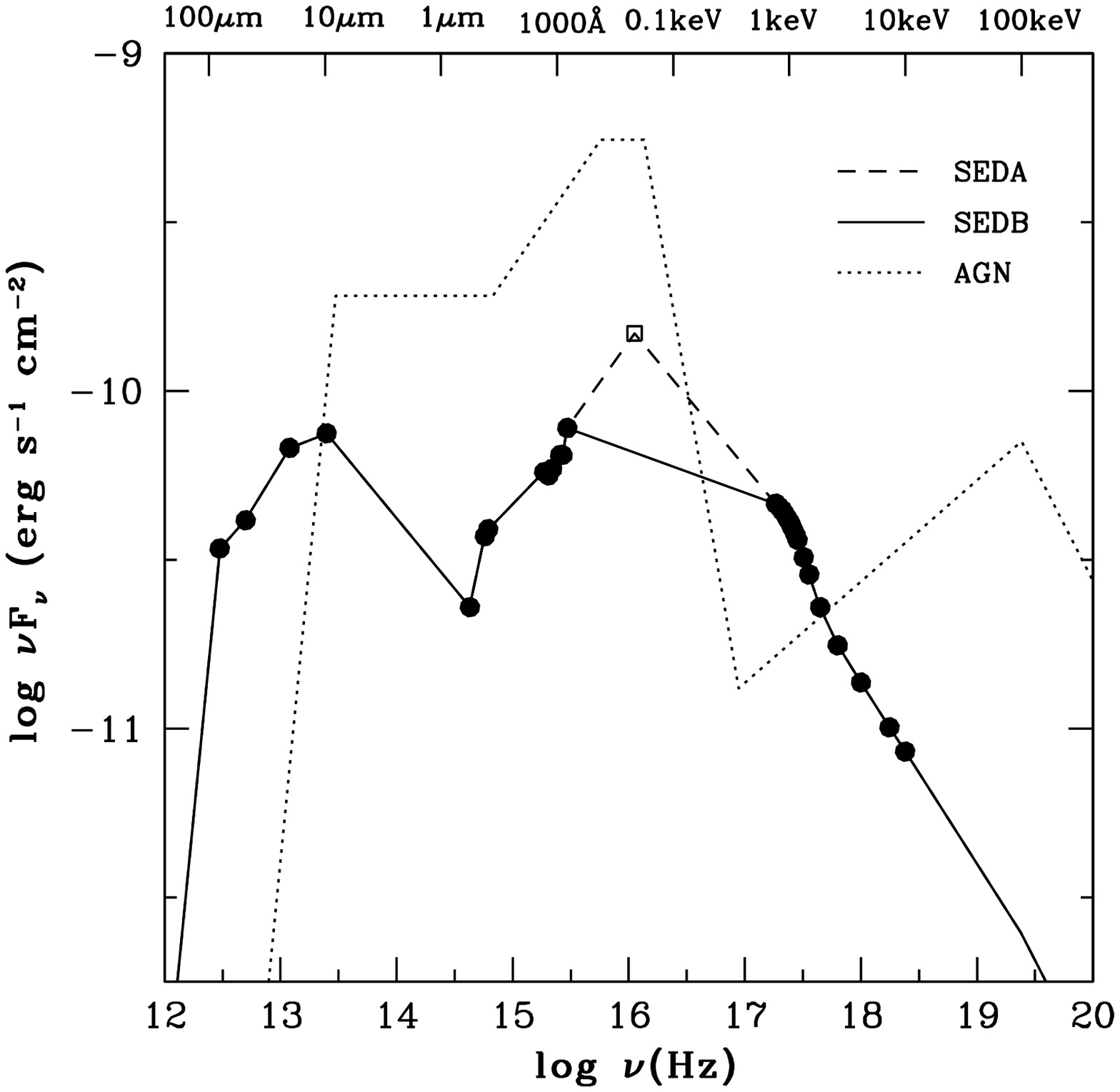}}

\centerline{}
\vspace{-1cm}

\figcaption{Input SEDs for {\tt Cloudy}, 
normalized to the absorption-corrected rest-frame flux at 2\,keV.
The dotted line is the ``table agn'' model in Cloudy; 
the short-dashed line is the SEDA described in \S\ref{sa:sed}
(the circles are the observed points, the empty square the extrapolated point); 
the solid line is the conservative SEDB described in \S\ref{sa:sed}. 
\label{sa:cloudyseds}}

\centerline{}
\vspace{-0.5cm}

\centerline{\includegraphics[width=10.5cm,height=9.5cm]{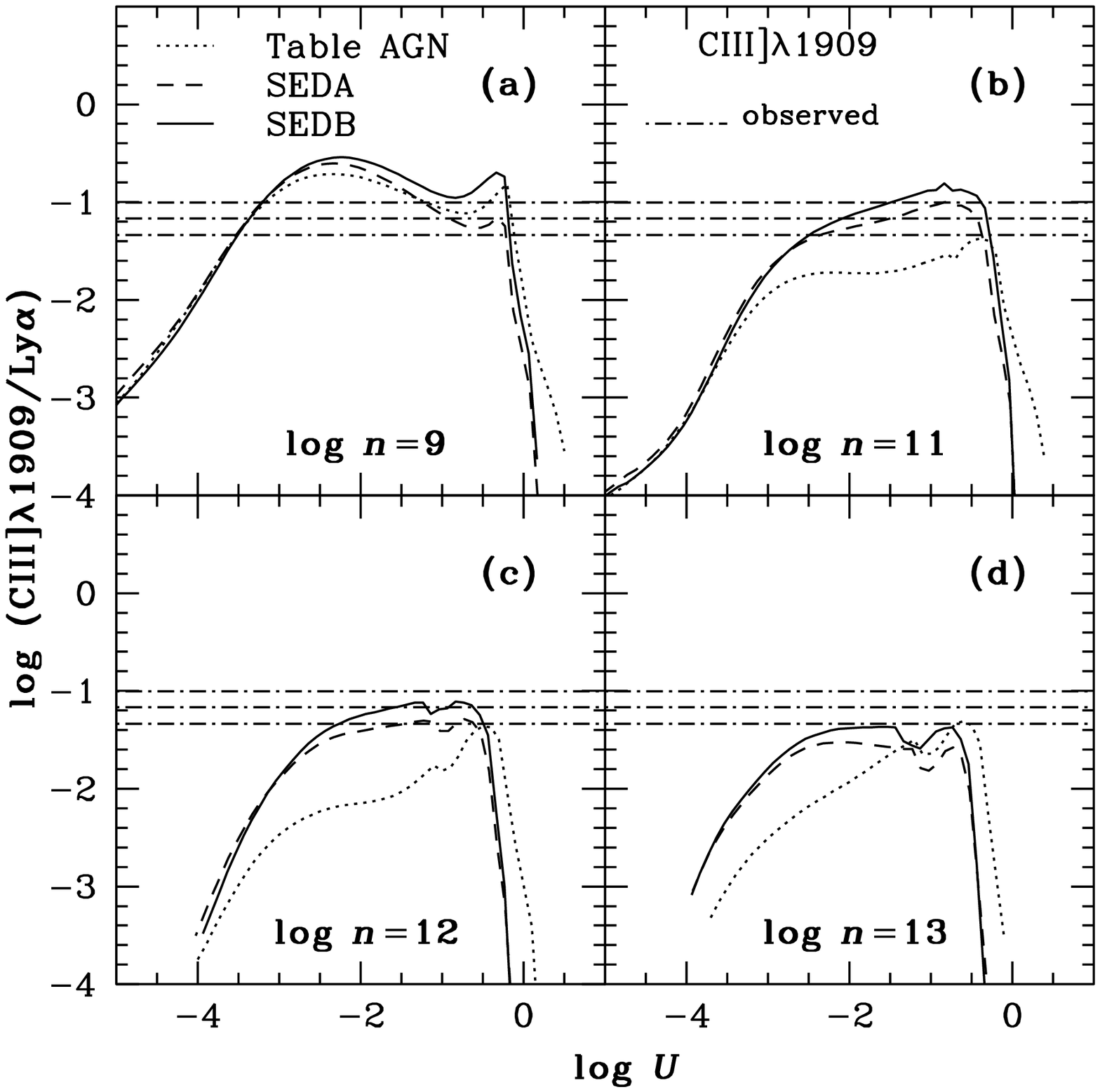}}

\centerline{}
\vspace{-0.5cm}

\figcaption{Line intensity relative to Ly$\alpha$ of \ion{C}{3}] $\lambda$1909 
as a function of hydrogen density (log $n = 9, 11, 12, 13$) and input continuum
(``table agn'', SEDA, and SEDB discussed in \S\ref{sa:pionizacontinua}). 
The horizontal dash-dot lines correspond to the observed values and their errors.  
\label{sa:diag1909}}

\centerline{}
\vspace{0.3cm}

	\subsection{Physical Conditions of the Emission-Line Gas 
			\label{sa:pionizaresults}}

Figure~\ref{sa:diag1909} shows the line intensity of \ion{C}{3}]$\lambda$1909 
relative to Ly$\alpha$ as a function of hydrogen density 
(log $n$ (cm$^{-3}$) $= 9, 11, 12, 13$) and input continuum 
(``table agn'', SEDA, and SEDB discussed in \S\ref{sa:pionizacontinua}). 
The horizontal lines correspond to the observed value of \ion{C}{3}]$\lambda$1909 and its
errors listed in Table~\ref{sa:elratios}. 
Solutions to $U$ (as also shown in Figures~\ref{sa:diagc3andn5} and 
\ref{sa:otherdiags}) are double valued, but we prefer higher 
values based on line widths, as discussed below. 
As expected for a semiforbidden transition, 
the \ion{C}{3}]$\lambda$1909 line becomes collisionally suppressed as 
the density increases, arguing for 
an upper limit for the density of $\log n \ltsim 12$. 
Analogously, Figure~\ref{sa:diagc3andn5} shows the intensity of \ion{C}{3}$\lambda$977 
and total \ion{N}{5}$\lambda$1240 which indicate $\log n > 9$. 
Furthermore, the observed \ion{C}{3}]$\lambda$1909 implies $\log U \approx -3.3$ or 
$\log U \approx -0.7$ for $\log n = 9$, both of which are much lower than the values 
required for \ion{O}{6}, for which we derive $\log U \approx -1.1$ or 
0.8\footnote{We find $\log U \approx -0.9$ or 0.8 if we do not 
deconvolve BEL and NEL.} 
(from a plot analogous to Figure~\ref{sa:otherdiags}b relative to $\log n = 9$), 
i.e., no consistent result can be found for low densities.

Considering $\log n \approx 11$ as a plausible estimate of the density, 
we can investigate the value of  the ionization parameter. 
This has been a difficult problem always, because with only optical/UV 
observations, multiple ionization states of a single element are not 
observed and so the values of $U$ are highly model dependent. 
With the multiwavelength, multimission observations of Ark~564, 
we now have observations of both \ion{C}{3} (with {\it FUSE}\,) and
\ion{C}{4} (with {\it HST}\,), so we can actually {\it measure} 
the ionization parameter. 
As shown in Figure~\ref{sa:otherdiags}a, the 
\ion{C}{3}$\lambda977$/\ion{C}{4}$\lambda1550$ ratio constrains the 
ionization parameter to 
$\log U =[-2.88,-0.22]$\footnote{$\log U =[-3.07,-0.36]$ 
if we do not deconvolve BEL and NEL.}.
Studies of reverberation mapping have shown that the BLR is stratified 
\citep{PetersonWandel99}, and so a single value of $U$ cannot possibly 
correspond to the entire BLR, but the range of $U$ determined above 
must be the dominant range of $U$ where \ion{C}{4} emission is produced.
The higher ionization lines, e.g. \ion{O}{6}$\lambda\lambda1032,1038$ and 
\ion{He}{2}$\lambda1640$ are likely to be produced closer in with 
higher ionization parameter. 
Indeed, as shown in Figure~\ref{sa:otherdiags}b, and 
Figure~\ref{sa:diagc3andn5}d, somewhat higher values of $U$ are preferred
for \ion{O}{6} ($\log U =[-1.52,-1.36]$ or 
$\log U =[0.61,0.87]$\footnote{$\log U =[-1.44,-1.23]$ or $\log U =[0.64,0.79]$
if we do not deconvolve BEL and NEL.}) 
and \ion{N}{5} ($\log U =[-1.43,0.4]$). 

Figure~\ref{sa:otherdiags}c shows the \ion{C}{4}$\lambda$1550 intensity ratio  
and implies that either $\log U$ $=$ $[-2.94,-2.66]$ or $\log U$ $=$ $[0.18,0.34]$. 
Figure~\ref{sa:otherdiags}d shows the intensity ratio 
\ion{C}{4}$\lambda$1240/\ion{N}{5}$\lambda$1550, for which we derive $\log U =[0.07,0.39]$. 
The comparison with the BLS1 NGC~5548 shows that \ion{N}{5} is stronger in Ark~564 by a 
factor of $\sim 2.3$, while \ion{C}{4} is
weaker in Ark~564 by a factor of $\sim 4.4$ 
(\S\ref{sa:fluxmeasur} and Table~\ref{sa:elratios}). 
Hence, the observed \ion{C}{4}/\ion{N}{5} ratio may be roughly one order 
of magnitude smaller in Ark~564, and the observed limits in 
Figure~\ref{sa:otherdiags}d probably reflect an overabundance of N (or C depletion). 

To discriminate between the low-$U$ ($\log U \approx -1.5$) and high-$U$ 
($\log U \approx 0$) solutions, we derive the distance of the BELR gas 
from $R_{\rm BELR} = (Q / 4\,\pi \,c\, n \, U)^{1/2} \approx 
1.63  \times 10^{16}\, (Q_{55}/U)^{1/2}\,n_{11}^{-1/2}$ cm, 
where $Q_{55}=Q/10^{55}$, 
assuming the above values of $U$ and 
$n$, and the photon luminosity $Q$ 
from \S\ref{sa:sed}.
For $\log n = 11$ and $\log U \approx -1.5$, the inferred distance of the BELR
is $R_{\rm BELR}^{SEDA} = 9.2 \times 10^{16}$ cm for $Q$(SEDA), 
and $R_{\rm BELR}^{SEDB} =6.5 \times 10^{16}$ cm for $Q$(SEDB). 
For $\log n = 11$ and $\log U \approx 0$,  
$R_{\rm BELR}^{SEDA} = 1.63 \times 10^{16}$ cm 
and $R_{\rm BELR}^{SEDB} = 1.15 \times 10^{16}$ cm. 
For a central mass of $M= $

\centerline{}
\vspace{-1cm}

\centerline{\includegraphics[width=10.5cm,height=10.0cm]{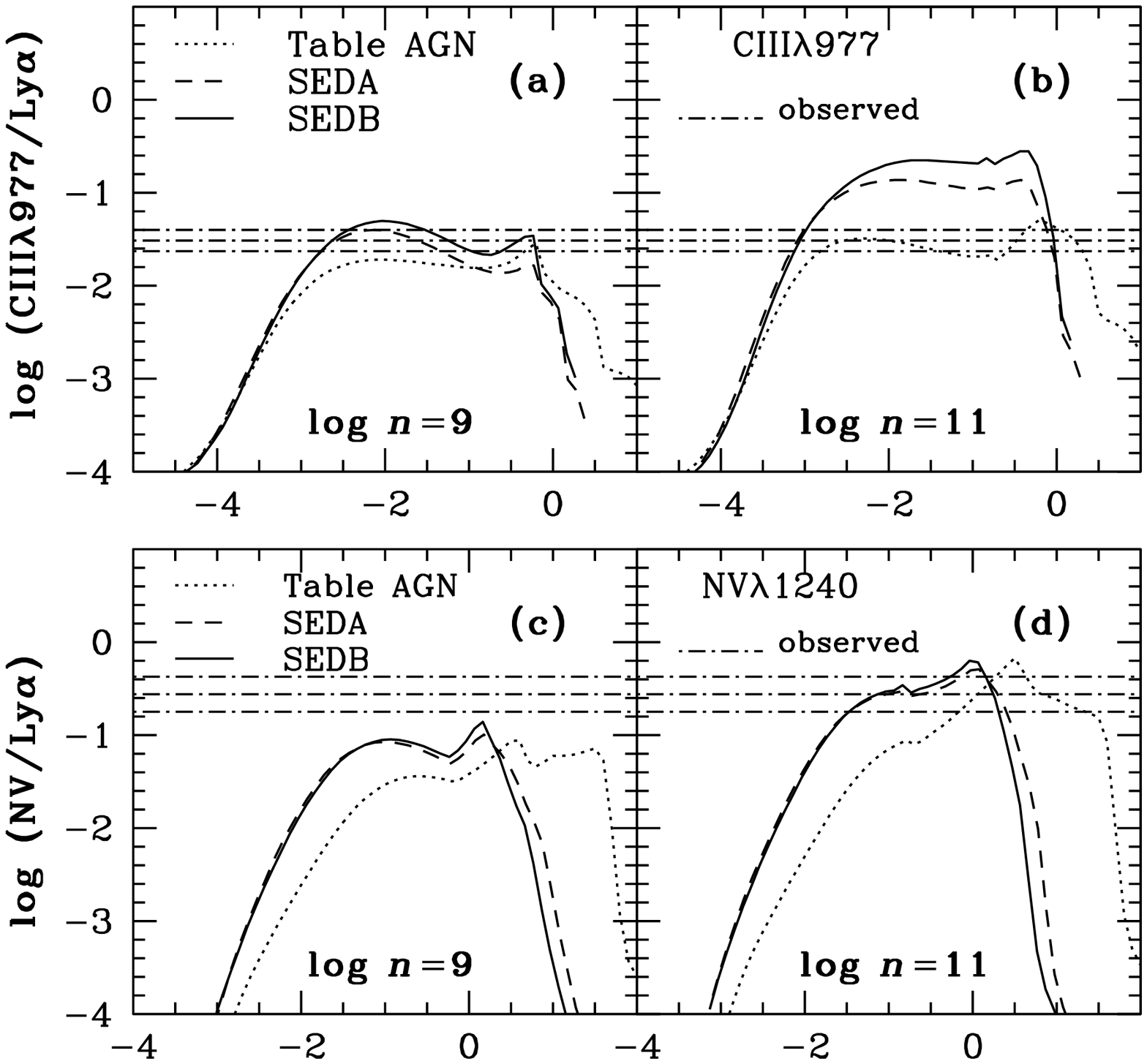}}
\figcaption{Same as 
Figure~\ref{sa:diag1909} for \ion{C}{3}$\lambda977$ (top) 
and \ion{N}{5}$\lambda1240$ (bottom) for log $n = $ 9 and 11. 
\label{sa:diagc3andn5}}

\centerline{\includegraphics[width=10.5cm,height=10.0cm]{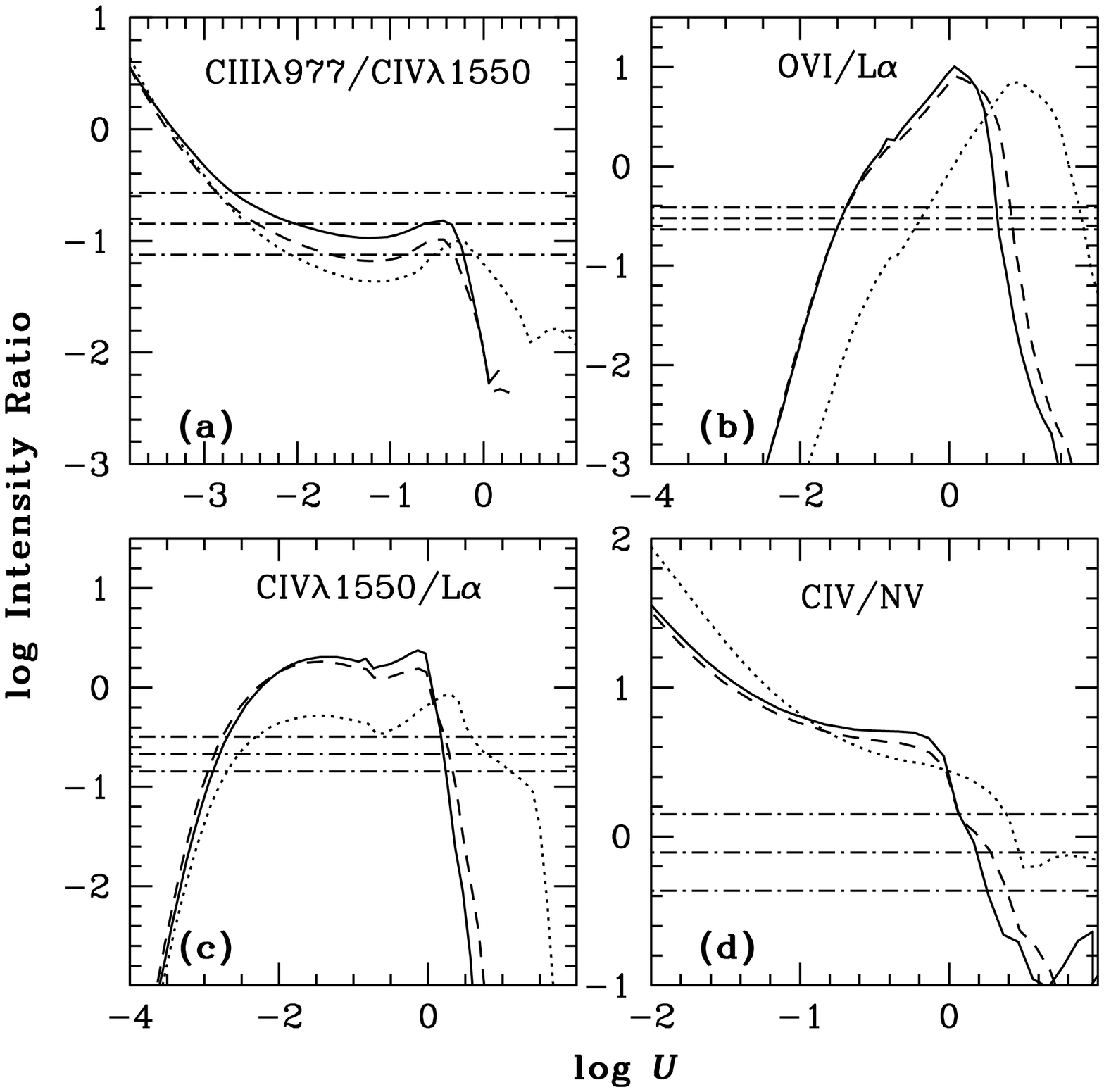}}
\figcaption{Line intensity
ratios of \ion{C}{3}$\lambda977 /$ \ion{C}{4}$\lambda1550$ (a),
\ion{O}{6}/Ly$\alpha$ (b), \ion{C}{4}$\lambda1550$/Ly$\alpha$ (c),
and \ion{C}{4}$\lambda1550 / $\ion{N}{5}$\lambda1240$ (d), as 
function of input continuum (same line notations as in Figure~\ref{sa:diag1909}).  
For all cases, log $n = 11$. 
\label{sa:otherdiags}}

\centerline{}
\vspace{0.3cm}

\noindent
$8 \times 10^{6} \Msun$, the expected velocity dispersion \citep{P2000} is $V = (GM/R_{\rm BLR}f)^{0.5}$, 
with $f=3/\sqrt{2}$. 
This corresponds to 740--880 \kms{} for the low-$U$ solutions 
and 1750--2090 \kms{} for the high-$U$ solutions.
Column (6) of Table~\ref{sa:elratios} shows that FWHM(\ion{C}{4})$=$ 1934 \kms, 
FWHM(\ion{C}{3})$=$ 1920 \kms, and 
FWHM(\ion{N}{5})$=$ 2809 \kms.  
Therefore, the comparison with the observed FWHMs favors the high-$U$ solutions.
Finally, Table~\ref{sa:elratios} shows that the FWHM of \ion{O}{6} is larger
than that of \ion{N}{5}, which is larger than that of \ion{C}{4}, 
again consistent with the stratified BLR model.

\noindent

	\section{Discussion\label{sa:discuss}}

A non-simultaneous optical, UV and X-ray spectral energy distribution 
of Ark~564 was presented 
by \citet{Comastriea01} who found that it peaks in the soft-X-ray band. 
Here we present a spectral energy distribution which is obtained from 
contemporaneous data covering almost 5 decades in energy\footnote{The 
Lick spectrum, which was obtained in 1980 is merely used to fill in
a small gap in the optical data, and the 2001 {\it FUSE} spectrum 
shows a continuum slope consistent with the one obtained from the 
{\it HST} spectrum. This latter fact suggests that, although 
the flux level changed between 2000 and 2001, the overall shape of 
the optical/FUV spectral energy distribution did not change. }. 
Simultaneity is particularly important for NLS1s, since, as a class, 
they are extremely variable in time, although Ark~564 has shown
only weak variability in the optical/UV bands (Paper II-III). 

We report some relevant data from the spectral energy distribution 
in Table~\ref{sa:sedvalues}. 
These were derived using the simple parameterization 
of the combination of an optical/FUV power law with spectral index 
$\alpha =0.42$ breaking 
at $\sim 50$\,eV to $\alpha_{ASCA} = 1.538$ (SEDA in Figure~\ref{sa:sed}),
as well as the {\it IRAS} and {\it IRTF} data points (\S\ref{sa:dataredux}).  
A more conservative spectral energy distribution, instead, connects 
the FUV and the soft X-ray data with a simple power law 
($\alpha = 1.08$, the solid line in Figure~\ref{sa:sedcorr}), 
and is a combination of an optical/FUV power law 
with $\alpha =0.42$ breaking at 1000\AA{} 
to $\alpha =1.08 1$, then again breaking at 
$\approx 0.8$\,keV to $\alpha_{ASCA} = 1.538$ (SEDB in Figure~\ref{sa:sed}). 
The ambiguity in the shape of the spectral energy distribution in the
900\AA--0.8\,keV region is rather unfortunate, since a considerable 
portion of the energy of Ark~564 might be output  in this range. 
Previous {\it ROSAT}  \citep{Brandtea94} 
and {\it BeppoSAX} \citep{Comastriea01}
data showed a flattening of the soft excess toward the lowest 
X-ray energies available,
and the {\it XMM} spectrum obtained during the monitoring campaign of 
2000 shows a definite curvature in the soft excess \citep{Vignaliea03}. 
 
An interesting issue is how Ark~564 compares to other NLS1 galaxies and 
with BLS1 in terms of its broad-band properties, as they can be quantified
by spectral indices. 
Table~\ref{sa:aox} reports the intrinsic spectral indices 
calculated between different wavelength bands for Ark~564 and, as a comparison, 
the corresponding values for the NLS1 Ton~S180 and BLS1s.
These show that while the inter-band properties of Ton~S180 are 
not significantly different from the ones observed in BLS1s \citep{Turnerea02}, 
this may not be the case for Ark~564. 
Table~\ref{sa:aox} indicates that the two NLS1s have steeper X-ray slopes than BLS1s, 
Ark~564 more so than Ton~S180 ($\alpha_{\rm x} = 1.57$ and 1.44, respectively,
compared to 0.91 for BLS1s), which is consistent with the general 
characteristics of NLS1s. 
The optical/X-ray spectral index ($\alpha_{\rm ox}$), on the other hand,
is lower in Ark~564 than in BLS1s and Ton~S180 with 
$\alpha_{\rm ox}=1.11$ for Ark~564\footnote{The observed $\alpha_{\rm ox}$ is 0.94 
in Ark~564.}, 1.52 for Ton~S180.
This reflects the fact that Ark~564 is relatively more X-ray bright, 
or that the optical continuum is suppressed, compared to other AGNs.
The other indices also reflect the X-ray brightness of Ark~564 when 
compared to BLS1s. 
We cannot exclude the possibility that intrinsic reddening in excess 
of the 

\centerline{}
\vspace{-1.5cm}

\centerline{\includegraphics[width=10.5cm,height=10.0cm]{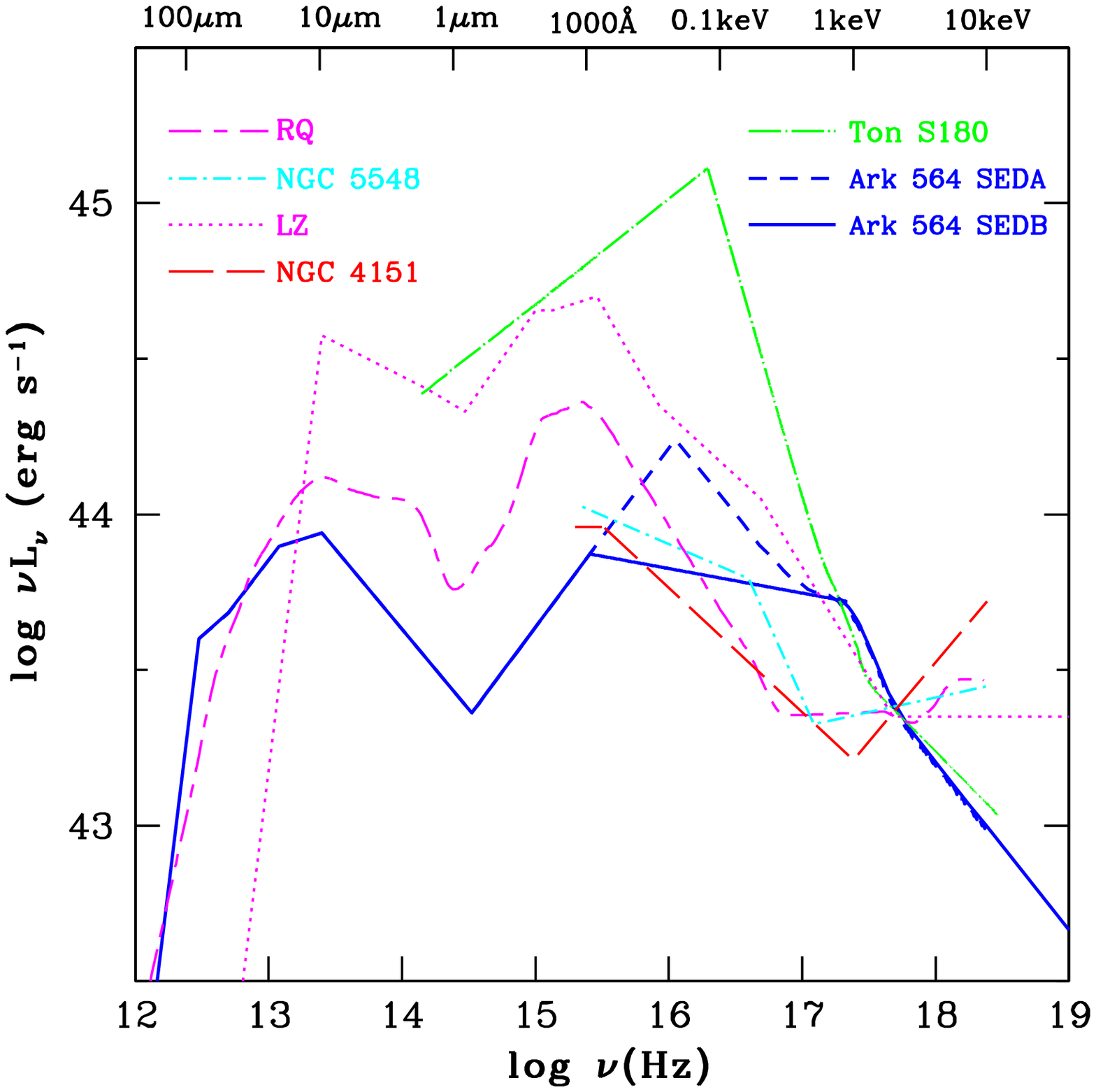}}

\figcaption{Comparison 
of the SEDs of Ark~564 with 
the mean SED for radio-quiet quasars (RQ; \citealt{Elvisea94}) 
and LZ \citep{Laorea97,Zhengea97}, 
one Seyfert 1 galaxy (NGC~4151,  \citealt{kraemerea00}),
one Seyfert 1.5 galaxy (NGC~5548, \citealt{KCFP98},
and the NLS1 Ton~S180 \citep{Turnerea02}. 
[{\it A color version of this plot is
available in the electronic edition of the Journal.}]
\label{sa:compseds}}

\centerline{}
\vspace{0.3cm}

\noindent 
one considered here may be present that could be suppressing the optical, 
but we argue that it would require the continuum to be more 
reddened than the emission lines 
(e.g.\ \ion{He}{2} $\lambda 1640/\lambda 4686$ ratio; Paper IV) show.
If these differences, however,  are not due to intrinsic reddening, then they 
represent a true difference in the energy balance of this NLS1 and 
that the X-ray emission is therefore consistent with the predictions of the 
slim disk models, indicating high accretion rates.

We estimated the luminosity in different energy bands using our 
parameterizations of the spectral energy distribution described above 
(SEDA and SEDB). Though with a $\sim 20$\% systematic error in the 
0.3--0.6\,keV band flux (C.\ Vignali 2003, private communication), 
the {\it XMM} spectrum provides a way to extend the energy range 
below the $\sim$ 0.8\,keV limit of the {\it ASCA} spectrum. Therefore, we 
also estimated the luminosity for the best-fit model to this spectrum. 
Table~\ref{sa:enbudget} reports these luminosities, 
and indicates that more energy is emitted in the 10--100\,eV band than 
in any other decade, constituting roughly half of the emitted energy in the
opt/X-ray ranges. This implies that the primary spectral component
peaks in the extreme UV--soft X-ray band and
this is generally consistent with disk-corona models \citep{hm91}. 
The integrated luminosity between $10^{-5}$ and 10\,keV 
gives a lower limit on the bolometric luminosity of 
$L_{\rm bol} \gtsim 10^{45}$\,\ergsec.

Growing evidence has been gathered that NLS1s 
are super-Eddington accretors. 
\citet{CH01} and  \citet{Collinea02} studied accretion rates in a 
sample of AGNs using the \citet{Kaspiea00} relationship, 
and found that not only half of their sample accretes close to or 
above the Eddington rate, but also that the largest Eddington ratios are 
found in NLS1s. Indeed, they showed that NLS1s are at the extreme of a 
well-defined sequence relating the Eddington ratio to the line widths.
Furthermore, \citet{WN03} show that a model of extreme slim disk 
(which is responsible for the soft X-ray excess, or hump,  seen in 
most NLS1s) and a hot corona (contributing to the hard X-ray emission) 
can also naturally explain the X-ray spectral variability  characteristics
observed in Ark~564 such as simultaneous variations of the soft hump 
and the hard X-ray without a significant time lag \citep{Akn564I}.
Through a comparison of the X-ray variability of Ark~564 and the 
BLS1 NGC~3516, \citet{Poundsea01} estimate that the mass of the central 
black hole in Ark~564 is $\sim 10^{7}\,\Msun$, implying an accretion rate 
in the range $\dot{m} \approx 0.2$--1. 
With a bolometric luminosity in the order of $10^{45}$\,\ergsec{} and an 
Eddington luminosity $L_{\rm Edd} \approx 10^{45}$\,\ergsec, we also infer
that $\dot{m} \approx 1$.
\citet{WN03} also provide a means of estimating the black hole mass
(if the accretion is super-critical) which is independent on the 
accretion rate itself. For Ark~564, we obtain 
$M \approx 2 \times 10^{6} \Msun$, which is within a factor of a 
few from the \citet{Collierea01} estimate. 

Figure~\ref{sa:compseds} compares the SED of Ark~564 with 
the mean SED for radio-quiet quasars \citep{Elvisea94}, 
the LZ SED \citep{Laorea97,Zhengea97}, 
the Seyfert 1.5 galaxies NGC~5548 \citep{KCFP98} and 
NGC~4151 \citep{kraemerea00}, 
and the NLS1 Ton~S180 \citep{Turnerea02}. 
There are significant differences in the intrinsic shape of the SED 
across the AGN population (see, also, \citealt{Turnerea02}), 
the most evident being the energy of the peak and the presence 
(or lack of) of the big blue bump (BBB), the signature of 
the emission from the accretion disk. 
The radial dependence of the temperature for an optically thick, 
geometrically thin accretion disk \citep{SS73} is,  
$T(R) \sim 6.3 \times 10^{5} \; (\dot{m})^{1/4} \; 
M_{8}^{-1/4} \; (R/R_{\rm S})^{-3/4} {\rm K}$ \citep{P2000}, 
where $M_8$ is the mass in units of $10^8 \Msun$, 
$R$ is the radius, and $R_{\rm S}$ is the Schwarzschild radius.
For Ark~564, using $\dot{m}\approx 1$, as we derived above, and 
$M \approx 8 \times 10^{6}\,\Msun$ (Paper~II), 
the peak temperature is $\sim 125$\,eV; 
this is within a factor of 3 from the peak of our less conservative 
parameterization, SEDA (defined extrapolating the optical power-law
continuum to meet the extrapolation of the X-ray power law) 
which peaks at 50\,eV. The true SED probably peaks somewhere 
between these two values. 
For comparison, the NLS1s RE~J1034$+$396 \citep{Puchnaea01}  
and Ton~S180 \citep{Turnerea02} 
peak at $\approx 250$\,eV and $\ltsim 100$\,eV, respectively.
Therefore, even among the NLS1s, 
differences in the shape of the SED are observed.
However, in none of these NLS1s there is an indication of the presence of 
optical/UV BBB, and  a strong soft X-ray excess is seen, instead. 
In this light, Ark~564, is also consistent with the paradigm 
that the accretion disk is so hot in NLS1s that the BBB is shifted 
in the EUV--soft X-rays. 

We also note that Ark~564 is rather FIR bright (with respect to the optical), 
compared to the sample of radio-quiet quasars and the LZ sample, as can be 
seen in Figure~\ref{sa:compsedsIR}, where the SEDs have been normalized to 
match their optical/UV slope 
(as opposed to the 2\,keV flux in Figure~\ref{sa:compseds}). 
Indeed, if we use the definition of the IR flux as a function 
of the {\it IRAS} fluxes given by 
\citet{SandersMirabel96}\footnote{$F(8-1000\mu$m$) = 
1.8 \times 10^{-14} \{13.48 f_{12} +5.16 f_{25} + 2.58 f_{60} 
+f_{100}\}$ W m$^{-2}$, where $f_{12}, f_{25}, f_{60}$, and  
$f_{100}$ are the {\it IRAS} flux densities in Jy 
at 12, 25, 60 and 100\,$\mu$m.}, we obtain that 
$L(8-1000\mu$ m$) \approx 10^{11} \Lsun$, which makes Ark~564 a 
luminous IR galaxy. The shape of the IR/optical SED of Ark~564 also 
resembles the shapes observed in the {\it IRAS} Bright Galaxy Survey
(see Figure~2 in \citealt{SandersMirabel96}, for intermediate values 
of $f_{60}$).
Using SEDA, we

\centerline{}
\vspace{-1cm}

\centerline{\includegraphics[width=10.5cm,height=10.0cm]{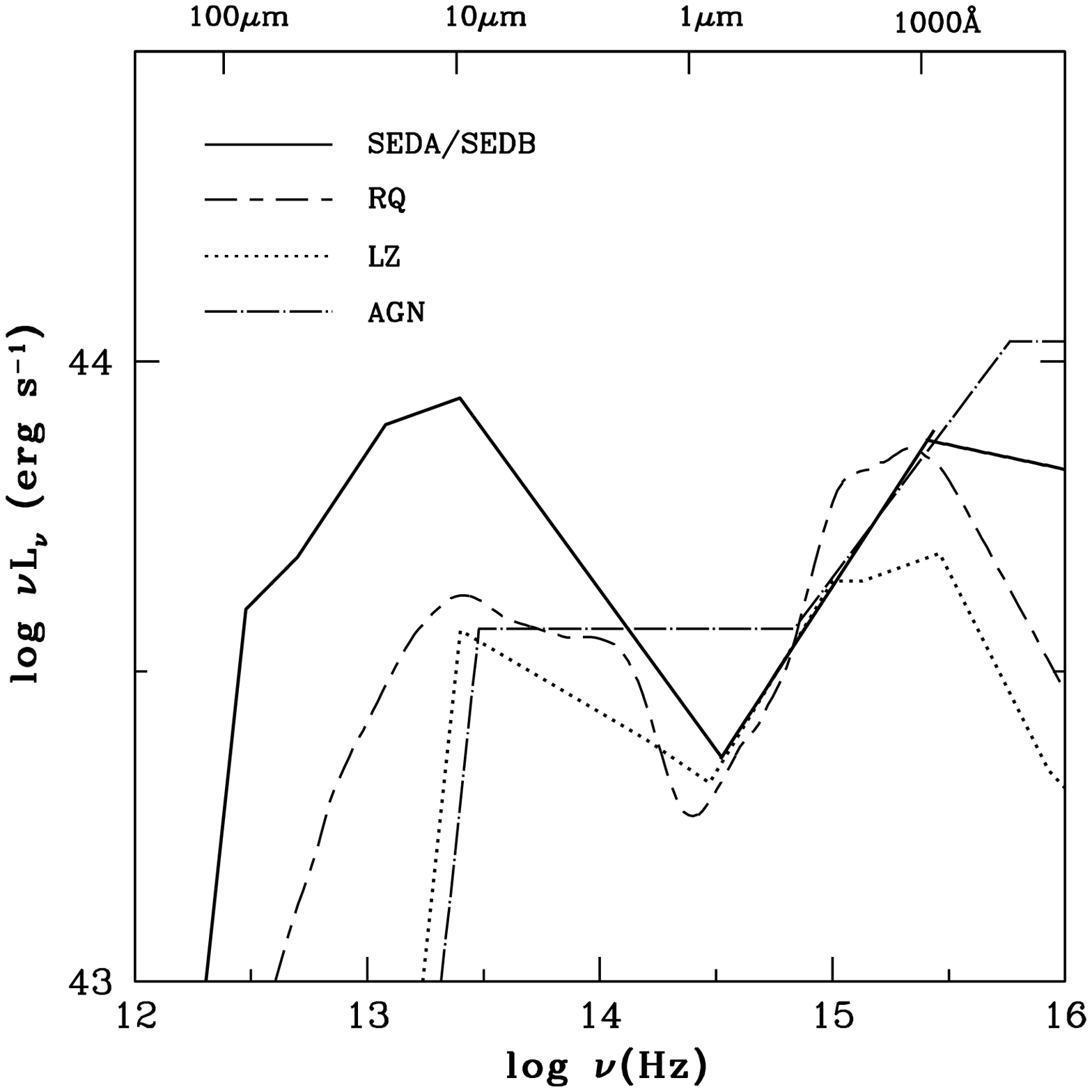}}

\centerline{}
\vspace{-0.3cm}

\figcaption{Comparison 
of the Ark~564 SED  with the mean SED for radio-quiet quasars (RQ; \citealt{Elvisea94}) ,
LZ \citep{Laorea97,Zhengea97}, and the ``table agn'' model in Cloudy (AGN). 
The SEDs are normalized
so that they match in the optical/UV part.
\label{sa:compsedsIR}}

\centerline{}
\vspace{0.3cm}

\noindent
estimate that the FIR luminosity is 
$L_{\rm FIR} \approx 1.3\times 10^{44}$\,\ergsec, i.e., 
$\sim 10\,\%$ of the total luminosity and $\sim 20\,\%$ of the combined
optical/UV/X-ray luminosity. 
\citet{Crenshawea02} noted that the associated warm UV absorber 
is lukewarm and dusty. The IR emission observed in this object could 
then be thermal emission from the dust grains embedded in the absorber
as they are heated by the strong UV/EUV continuum. 
Given the IR brightness in this object and in many NLS1s 
\citep{Moranea96}, it is not unlikely that a contribution might 
be coming from the host galaxy, in the form of a nuclear starburst
\citep{Mathur00}. 
\citet{Crenshawea02}, however, did not detect any extended emission 
in the two-dimensional STIS spectral images that could be due to 
a nuclear starburst. We compared the {\it FUSE} spectrum, which was taken
through a much larger aperture, and hence is more likely to show 
stellar absorption features, with a {\it FUSE} `template' spectrum of 
the starburst galaxy NGC~7496; using the constraints from the 
Ly$\gamma$--\ion{C}{3}$] \lambda$977 profiles, 
we obtained a rough upper limit of the 
starburst contribution at 1000\AA{} of 50\%.
A further constraint on the IR starburst contribution would probably come 
from detection and measurements of the 3.3--3.4$\mu$m PAH emission features, 
which have been found in starburst galaxies, luminous IR galaxies, 
and obscured AGNs \citep{Moorwood86,Imanishi02}, and which have been 
successfully detected in the NLS1 NGC~4051 \citep{rodriguezviegas03}.

Using our SEDA and SEDB as inputs to {\tt Cloudy} we predicted 
the intensity of the strongest lines in the FUV/UV spectrum 
and compared them with the observed values in order to constrain 
the physical parameters of the line-emitting gas, namely, the density 
and ionization parameter. 
Figure~\ref{sa:diagc3andn5} and \ref{sa:otherdiags} show 
that SEDA and SEDB, because of their strong
EUV to soft X-ray flux, for a given observed line ratio 
predict values of ionization parameter 
which are lower than those with standard AGN continuum. 
From  \ion{C}{3}] $\lambda$1909, \ion{C}{3} $\lambda$977, 
and \ion{N}{5} $\lambda$1240, we infer that $\log n \approx 11$.
Two classes of solutions for $U$ are consistent with this density value, one with 
low $U$ values ($\log U \approx -1.5$) and one with high $U$ values ($\log U \approx 0$).
We discarded the low-$U$ class (\S\ref{sa:pionizaresults}) on the basis that 
the predicted widths of the lines, derived from the velocity dispersion  
$V = (GM/R_{\rm BLR}f)^{0.5}$, of 740--880 \kms{} are too small with 
respect to the observed ones ($\approx 2000$ \kms; Paper~II). 
As expected, we find that the BLR is stratified around $\log U \approx 0$, 
with higher ionization lines originating from regions with higher $U$.

Column (6) of Table~\ref{sa:elratios} shows that FWHM(Ly$\alpha$)$=$ 2114 \kms, 
FWHM(\ion{N}{5})$=$ 2809 \kms, and FWHM(\ion{C}{4})$=$ 1934 \kms, 
which indicate that the radii of the L$\alpha$, \ion{N}{5}, and \ion{C}{4} 
broad-line emitting regions are 
$R_{\rm BLR}^{{\rm L}\alpha} \approx 4.3$ lt-days, 
$R_{\rm BLR}^{\rm N \sc V} \approx 2.5$ lt-days, 
and $R_{\rm BLR}^{\rm C \sc IV } \approx 5.2$ lt-days. 
Using the findings of previous monitoring programs on Seyfert 1s 
\citep{NetzerPeterson97}, we can estimate the size of the H$\beta$-emitting region
from $R_{\rm BLR}^{{\rm L}\alpha} \approx 0.5 R_{\rm BLR}^{{\rm H}\beta}$, and
$R_{\rm BLR}^{\rm N \sc V} \approx 0.2 R_{\rm BLR}^{{\rm H}\beta}$. 
For NGC~5548, furthermore,  
$R_{\rm BLR}^{\rm C \sc IV} \approx 0.5 R_{\rm BLR}^{{\rm H}\beta}$ 
\citep{Peterson93}. 
We can conclude that $R_{\rm BLR}^{{\rm H}\beta} \approx 10 \pm 2$ lt-days, which 
is consistent with the $R_{\rm BLR}^{{\rm H}\beta}$--luminosity 
relationships of \citet{Kaspiea00} and \citet{P2000}, when we assume a luminosity 
$\lambda L_{\lambda} (5100$ \AA)$ \approx 3.2 \times 10^{43}$ \ergsec{} 
(Table~\ref{sa:sedvalues}). 
This indicates that the BLR radius of this NLS1 is consistent with the 
distribution of BLR radius in BLS1s, and that the narrowness of the emission lines 
is not due to the BLR being relatively further away from the central 
mass than in BLS1s of comparable luminosity. 

Table~\ref{sa:elratios} shows that some interesting 
differences in line ratios can be found with respect to the Sy1.5 NGC~5548. 
Indeed, \ion{N}{5} $\lambda$1240 is stronger in Ark~564 by a factor 
of $\sim 2.3$, while \ion{C}{4} $\lambda$1550, \ion{C}{3}] $\lambda$1909, 
and Mg II $\lambda$2800 are weaker in Ark~564 by a factor of $\sim 4.4$, 
2.5 and 2.8, respectively. 
While all line ratios in Ark~564 are statistically consistent with the ones 
measured for a mean QSO (given the large uncertainties on our measurements), 
Carbon lines are at the lower end and Nitrogen at the upper end of the 
QSO distribution, confirming this trend for weak Carbon and strong Nitrogen 
in Ark~564. 
Furthermore, \ion{C}{3}] $\lambda$977 would indicate that 
$-3.13 < \log U < -0.05$
(Figure~\ref{sa:diagc3andn5}), and for this range, 
the observed \ion{N}{5}$/$\ion{C}{4} ratio is larger
than the model predictions by a factor of $\sim 8$. 
This may imply 
super-solar metallicity in this NLS1 as suggested by \citet{Mathur00} and 
is consistent with the finding of \citet{Shemmerea02} that 
NLS1s have higher metallicities than BL AGNs for a given luminosity.

An interesting question is how sensitive the emission lines are to the
true shape of the ionizing continuum. Given the difficulty related to 
deblending the BEL and NEL components of the lines and consequent 
large errors on the observed line ratios, the present emission line data 
do not allow us to discriminate between SEDA and SEDB. 
These SEDs differ in the range 1000\AA--750\,eV (corresponding to the 
gap in the data between the {\it FUSE} and {\it ASCA} spectra), 
with the maximum difference at around 50\,eV. 
This difference should show the most for the high ionization lines of
\ion{C}{4}, \ion{N}{5} and \ion{O}{6}. 
However, the predicted strength of their {\it emission} lines is
very similar for the two SEDs, for a wide range of ionization parameters
of interest (Figures~\ref{sa:diagc3andn5} and \ref{sa:otherdiags}).
Perhaps, this is the reason why the emission line spectra of most AGNs 
look so very similar, over a wide range of luminosities.
This demonstrates how unsuitable emission lines are as diagnostics for
the underlying SEDs. 

The absorption lines, on the other hand, are sensitive to
the 

\centerline{}
\vspace{-1cm}

\centerline{\includegraphics[width=10.5cm,height=10.0cm]{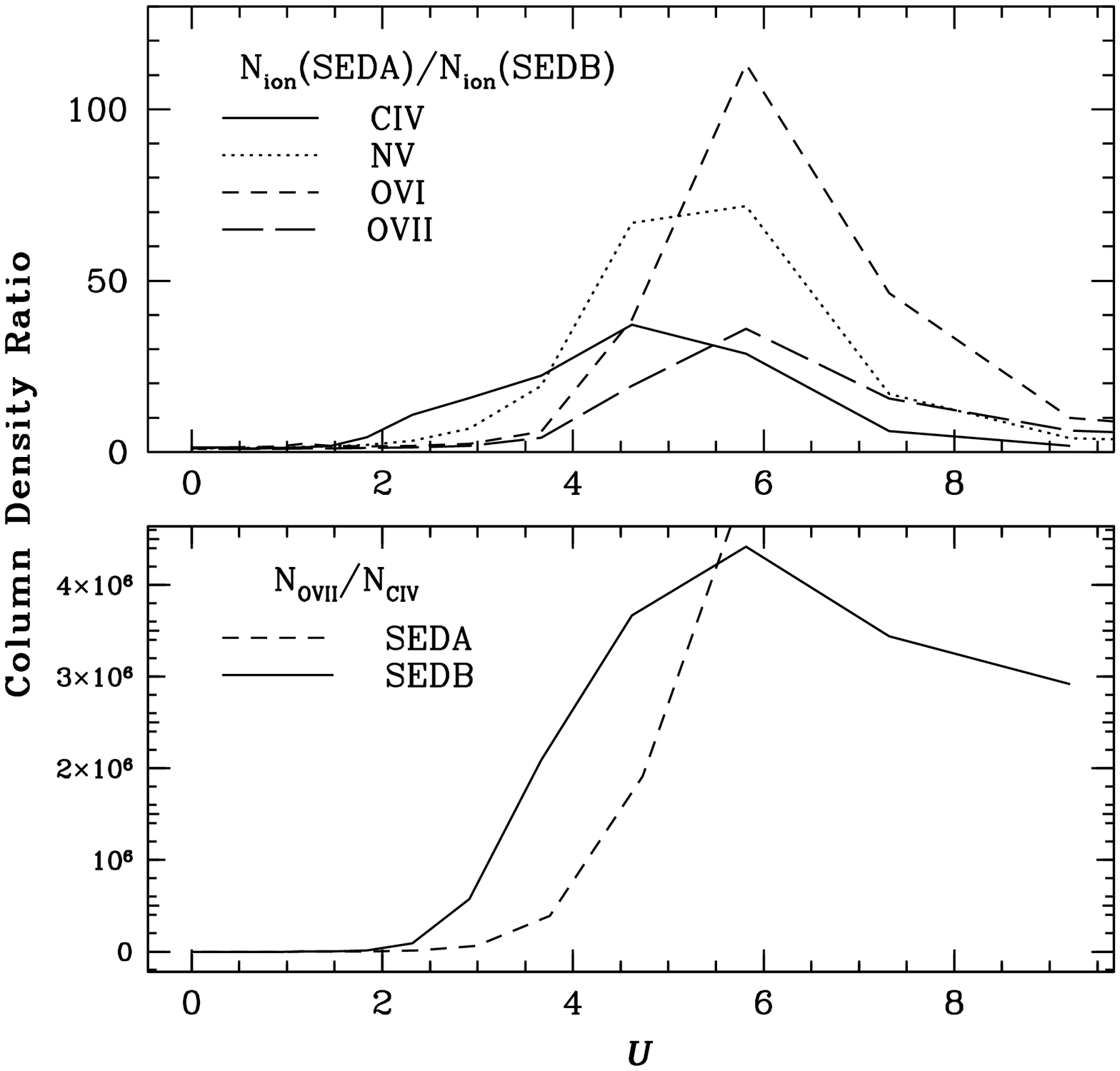}}

\centerline{}
\vspace{-0.3cm}

\figcaption{Top panel shows the ratio of the 
column densities relative to SEDA and SEDB as a function of $U$ 
for \ion{C}{4}, \ion{N}{5}, \ion{O}{6}, and \ion{O}{7}. 
Bottom panel shows the ratio of the column densities of \ion{O}{7}
to \ion{C}{4} as a function of $U$ for solar abundances, for SEDA and SEDB.
\label{sa:fions}
}

\centerline{}
\vspace{0.3cm}

\noindent
input SED \citep{MWEF94}. 
Column densities of different ions can be inferred from the 
the fractional abundances $f_{\rm ion}$,  
the total column density $N_{\rm H}$, 
and the assumed abundances $N_{\rm X}$,
$N_{\rm ion} = N_{\rm H}\, N_{\rm X}\,N_{\rm ion}$, 
through photoionization calculations (assuming an SED).
Figure~\ref{sa:fions} (Top) shows the fractional column densities for 
the same ion calculated with the different SEDs 
[$N_{\rm CIV}$(SEDA) /$N_{\rm CIV}$(SEDB), 
$N_{\rm NV}$(SEDA) /$N_{\rm NV}$(SEDB), 
$N_{\rm OVI}$(SEDA) /$N_{\rm OVI}$(SEDB), 
and $N_{\rm OVII}$(SEDA) /$N_{\rm OVII}$(SEDB)], 
and illustrates how the  predicted column densities of 
high ionization lines differ, 
with \ion{O}{6} showing the largest variations, 
with over two orders of magnitude difference.
Figure~\ref{sa:fions} (Bottom) compares the column 
density for \ion{O}{7} and \ion{C}{4} 
calculated with the same SED and solar abundances 
[$N_{\rm OVII}$(SEDA) / $N_{\rm CIV}$(SEDA) and
$N_{\rm OVII}$(SEDB)  / $N_{\rm CIV}$(SEDB)],
which turn out to be good diagnostics even at lower 
ionization parameters. 
Thus, in principle,
absorption line studies offer a far more powerful tool to determine the
ionizing continuum of AGNs, compared to the emission lines. This also
underscores the power of multiwavelength observations, as the maximum
discriminator comes from comparing the lower- and higher-ionization lines.

	\section{Summary\label{sa:summary}}

We presented the intrinsic spectral energy distribution of Ark~564, 
constructed with quasi-simultaneous data obtained during 
2000 and 2001. We compared this SED with that of Ton~S180 
and with those obtained for Broad-Line Seyfert 1s to infer 
how the relative accretion rates vary among the Seyfert 1 population. 
The peak of the SED is not well constrained; however, 
in our parameterization most of the energy of this object is 
emitted in the 10--100\,eV regime, and constitutes roughly half 
of the emitted energy in the optical/X-ray ranges. 
This is consistent with a primary spectral component 
peaking in the soft X-ray band, therefore with the predictions of the 
slim disk models, hence high accretion rates. Indeed, we  
estimate that  $\dot{m} \approx 1$.

We constrained the mean physical conditions in the BELR of this AGN,
by examining the emission lines observed in its spectrum, 
and deduced the physical properties of the line-emitting gas 
through photoionization modeling. We concluded that the line-emitting 
gas is characterized by $\log n \approx 11$ and $\log U \approx 0$,
and is stratified around $\log U \approx 0$.
Our estimate of the radius of the H$\beta$ emitting region 
$R_{\rm BLR}^{{\rm H}\beta} \approx 10 \pm 2$ lt-days, 
is consistent with the $R_{\rm BLR}^{{\rm H}\beta}$-luminosity 
relationships of \citet{Kaspiea00} and \citet{P2000}. 
This indicates that the narrowness of the emission lines is not 
due to the BLR being relatively further away from the central 
mass than in BLS1s of comparable luminosity. 
We also find evidence for super-solar metallicity in this NLS1, 
based on the low \ion{C}{4}/\ion{N}{5} observed line ratio.  
While the emission lines turn out to be unsuitable as diagnostics for
the underlying SEDs, we showed that absorption line studies 
offer a far more powerful tool to determine the
ionizing continuum of AGNs, especially if comparing 
the lower- and higher-ionization lines.
This underlines the power of multiwavelength observations.

\acknowledgements 

We thank C.\ Vignali for providing us with the {\it XMM} 
spectrum ahead of publication and the anonymous referee for 
suggestions that improved the paper. 
P.R. and S.M. acknowledge support through NASA grant NAG5-10320 
({\it FUSE}\,), and the Italian MIUR.
T.\ J.\ T. and  W.\ N.\ B.\  acknowledge support through 
NASA LTSA grants NAG5-7385 and NAG5-13035, respectively.
We also acknowledge support from HST--GO--08265.01--A from the 
Space Science Telescope Institute, which is operated by the Association of 
Universities for Research in Astronomy, Inc., under NASA contract 
NSS5-226555. 
This research has made use of the NASA/IPAC Extragalactic Database 
(NED) which is operated by the Jet Propulsion Laboratory, California 
Institute of Technology, under contract with the National Aeronautics 
and Space Administration.

\begin{deluxetable}{lllllc}	
 \tablewidth{0pc} 
 \tablecaption{Observing Log for Arakelian~564\label{sa:obslogs}}
\tablehead{\colhead{Observatory} & \colhead{Instrument} & \colhead{UT Dates} 
				      & \colhead{Wavelength/Energy\tablenotemark{a}} & \colhead{Notes} & \colhead{References} \\
	   \colhead{(1)} & \colhead{(2)} & \colhead{(3)} & \colhead{(4)} & \colhead{(5)} 
	& \colhead{(6)}}
\startdata 
{\it ASCA}     	& 	     & 2000 Jun 1--Jul 6  & 0.75--9.76\,keV & 2.98\,Ms, continuous\tablenotemark{b} 	  & 1 \\
{\it XMM-Newton} & 	     & 2000 Jun 17        & 0.3--8\,keV     & 	  & 2 \\
{\it FUSE}     	& 	     & 2001 Jun 29-30     & 1000--1175\,\AA  & 63\,ks; 30\arcsec x 30\arcsec (LWRS) & 3 \\
{\it HST}       & STIS/G140L & 2000 May 9--Jul 8  & 1175--1711\,\AA  & 554304\,s; 52\arcsec x 0\farcs5 	  & 4,5\\
{\it HST}       & STIS/G230L & 2000 May 9--Jul 8  & 1711--3143\,\AA  & 24216\,s; 52\arcsec x 0\farcs5 	  & 4,5\\
Lick 		& 	     & 1980		  & 3170--4160\,\AA  &   & 6 \\		
Wise 		& FOSC	     & 1998 Nov--2001 Jan & 4160--7790\,\AA  & 		  & 7\\
{\it IRTF}      & SPEX	     & 2000 Oct 11, 13 	  & 8200--24000\,\AA\tablenotemark{c}  & $\sim 30$\,min, 15\arcsec x 0\farcs8	  & 8 \\
{\it IRAS}      & 	     &          	  & 12, 25, 60, 100\,$\mu$m &	  &9 \\
\enddata 
\tablenotetext{a}{Observed-frame wavelength/energy bands utilized.}
\tablenotetext{b}{Except for gaps due to Earth 
occultation and passage of the spacecraft through the SAA.}
\tablenotetext{c}{Only few continuum points were used for this work.} 
\tablerefs{(1)  \citealt{Akn564I}. (2) \citealt{Vignaliea03}. (3) \citealt{Romanoea02b}.
(4) \citealt{Collierea01}. (5) \citealt{Crenshawea02}. (6) D.\ E.\ Osterbrock 2002, private communication; 
(7) \citealt{Shemmerea01}. (8)  \citealt{rodriguezea02b} and references therein. (9)  \citealt{iras}. }
\end{deluxetable}

\begin{deluxetable}{lcc}	
\tablewidth{0pc}
\tablecaption{Data from the Spectral Energy Distribution \label{sa:sedvalues}}    
\tablehead{\colhead{Rest Wavelength}  & \colhead{$\nu L_{\nu}$ (Observed)}  
	& \colhead{$\nu L_{\nu}$ (Intrinsic, SEDA)} \\
	   \colhead{/Energy} & \colhead{($\times 10^{43}$\,\ergsec)}  
	& \colhead{($\times 10^{43}$\,\ergsec)}	\\
  	   \colhead{(1)} & \colhead{(2)} & \colhead{(3)} 
}
\startdata
97.59\,$\mu$m\tablenotemark{a}		& 3.98	& 4.57\\ 
58.56\,$\mu$m\tablenotemark{a}		& 4.82	& 5.50\\
24.40\,$\mu$m\tablenotemark{a}		& 7.91	& 8.91\\
11.71\,$\mu$m\tablenotemark{a}		& 8.73	& 9.77\\
2.4\,$\mu$m\tablenotemark{b}		& 3.3	& 3.5 \\
1.6\,$\mu$m\tablenotemark{b}		& 2.2	& 2.4 \\
1.14\,$\mu$m\tablenotemark{b}		& 2.1	& 2.4 \\
1\,$\mu$m\tablenotemark{c} 		& 2.045   & 2.165 \\	
9850,{\rm \AA}\tablenotemark{b}         & 2.0  & 2.3 \\
7000\,{\rm \AA}           		& 1.851  & 2.660 \\
5500\,{\rm \AA}           		& 1.73   & 3.057 \\
5100\,{\rm \AA}           		& 1.694   & 3.193 \\
3000\,{\rm \AA}           		& 1.461  & 4.338 \\
2500\,{\rm \AA}           		& 1.388  & 4.819 \\
1000\,{\rm \AA}           		& 1.075  & 8.178\\
0.046\,keV\tablenotemark{d}		& \nodata & 17.5 \\
0.25\,keV\tablenotemark{d}		& \nodata & 7.195 \\
0.78\,keV				& 2.895 & 5.240 \\
1\,keV                    		& 3.046 & 4.668   \\
2\,keV                    		& 2.060 & 2.419  \\
10\,keV                   		& 0.942	& 0.970  \\ 
\enddata
\tablecomments{The intrinsic optical/UV/X-ray data are from the 
reddening-corrected, rest-frame SEDA (\S\ref{sa:sed}). 
SEDB is SEDA with the exclusion of the point at $\sim50$\,eV. 
We adopt $H_0=75$ ${\rm km\ s ^{-1}\ Mpc^{-1},}$  $q_0=0.5$.} 
\tablenotetext{a}{{\it IRAS} data points (\S\ref{sa:dataredux}).}
\tablenotetext{b}{{\it IRTF} data points (\S\ref{sa:dataredux}).}
\tablenotetext{c}{Extrapolated value from the optical/FUV power law 
				(spectral index $\alpha =0.42\pm 0.01$, \S\ref{sa:sed}).}
\tablenotetext{d}{Extrapolated value from the {\it ASCA} power law 
		($\alpha_{ASCA} =1.538\pm 0.005$, \S\ref{sa:mainres}), 
		peak of SEDA (\S\ref{sa:sed}).}
\end{deluxetable}  
\clearpage

\begin{deluxetable}{lcccccc}
\tablewidth{0pc}
\tablecaption{Arakelian~564 Emission-Line Characteristics \label{sa:elratios}}
\tablehead{
\colhead{Line} 	& \colhead{F$/$F(Ly$\alpha$)} & \colhead{F$/$F(Ly$\alpha$)} 
		& \colhead{F$/$F(Ly$\alpha$)} & \colhead{F$/$F(Ly$\alpha$)} 
		& \colhead{FWHM}  & \colhead{References}  \\
\colhead{} 	& \colhead{BEL+NEL\tablenotemark{a}} & \colhead{BEL\tablenotemark{a}} 
	     	& \colhead{QSO\tablenotemark{b}} &  \colhead{NGC5548\tablenotemark{c}} 
		& \colhead{(\kms)\tablenotemark{d}} & \colhead{} \\
\colhead{(1)} & \colhead{(2)} & \colhead{(3)} & \colhead{(4)} & \colhead{(5)} 
		& \colhead{(6)} & \colhead{(7)}
}
\startdata
\ion{C}{3} $\lambda$977      	& 0.053$\pm$0.017 &0.031$\pm$0.015 	&\nodata    				&\nodata 	&4000,1100 	&1,2 \\
\ion{O}{6} $\lambda$1032  	& 0.284$\pm$0.032 &0.200$\pm$0.052 	&\nodata    		 		&0.036 	&4000,1100	&1,2 \\
\ion{O}{6} $\lambda$1038  	& 0.141$\pm$0.048 &0.100$\pm$0.026 	&\nodata    		 		&0.018 	&4000,1100	&1,2 \\
\ion{O}{6} $\lambda$1032,1038  	& 0.425$\pm$0.141 &0.300$\pm$0.076 	& 0.1--0.3\tablenotemark{e} 		&0.054 	&4000,1100	&1,2 \\
L$\alpha$ $\lambda$1216   	& 1.00  	  & 1.00 	   	& 1.00      		  		& 1.00  	&2114 		&3,4 \\
\ion{N}{5} $\lambda$1240  	& 0.275$\pm$0.118 &\nodata 		& 0.09--0.26 		    		& 0.119 	&2809 		&3,4 \\
\ion{C}{4} $\lambda$1550  	& 0.215$\pm$0.087 &\nodata 		& 0.28--0.42 		    		& 0.937 	&1934		&3,4 \\
\ion{He}{2} $\lambda$1640       & 0.101$\pm$0.040 &\nodata 		& 0.06--0.13\tablenotemark{f} 		& 0.143\tablenotemark{f} &1195,1831  &3,4 \\
\ion{O}{3}] $\lambda$1663      	& 0.038$\pm$0.015 &\nodata 		&  \nodata 		 		& \nodata	&\nodata	&3,4 \\
\ion{C}{3}] $\lambda$1909      	& 0.068$\pm$0.026\tablenotemark{g}&\nodata &0.09--0.19\tablenotemark{h}		& 0.171\tablenotemark{g} &1920 	&3,4 \\
Mg II $\lambda$2800       	& 0.066$\pm$0.008 &\nodata 		& 0.06--0.13 		 		& 0.188 	&1659 		&3,5 \\
H$\beta$ $\lambda$4861    	& 0.063$\pm$0.019 &\nodata 		& 0.02--0.05 		    		& \nodata 	&700  	&3 \\
H$\alpha$ $\lambda$6563   	& 0.240$\pm$0.080 &\nodata 		& 0.05--0.09\tablenotemark{i}		& \nodata	&\nodata 	&3
\enddata
\tablenotetext{a}{Reddening-corrected flux relative to Ly$\alpha$ 
derived from \citet{Crenshawea02}. 
The lines are corrected using $E(B-V)$ $=$ 0.14 $\pm$ 0.04 mag and 
Ark~564 reddening curve from \citet{Crenshawea02} 
plus $E(B-V)$ $=$ 0.03 mag and Galactic curve (\S\ref{sa:reddcorr}).}
\tablenotetext{b}{Data from the mean observed QSO spectrum in \citet{Baldwinea95} and references therein,
			corrected for the reddening appropriate for Ark~564.}
\tablenotetext{c}{BEL fluxes, corrected for NEL contribution and Galactic reddening of the Sy1.5 NGC~5548;
	 derived from \citet{KoristaGoad00} and references therein. 
	The \ion{O}{6} line ratios are derived from the BEL values in \citet{Brothertonea02},
	then corrected for Galactic reddening ($E(B-V)$ $=$ 0.03 mag, 
	and extinction law of \citealt{card89}).}
\tablenotetext{d}{Model FWHM of \ion{C}{3}$\lambda 977$ and \ion{O}{6} 
			relative to BEL and NEL components (Paper~V), separately; 
		   the others are measured on the whole line profile (Paper~II). 
		   The \ion{He}{2} values are relative to the G140L and 
			G230L mean spectrum, respectively  (Paper~II).}
\tablenotetext{e}{Total \ion{O}{6}$+$Ly$\beta$ flux. This compares with $0.434\pm0.141$ for Ark~564.}
\tablenotetext{f}{Total \ion{He}{2}$+$\ion{O}{3}]$\lambda$1666 flux.
				This compares with $0.139\pm0.043$ for Ark~564.}
\tablenotetext{g}{Total \ion{C}{3}] $\lambda$1909$+$\ion{Si}{3}] $\lambda$1892 flux.}
\tablenotetext{h}{Total \ion{C}{3}] $\lambda$1909$+$\ion{Si}{3}] $\lambda$1892$+$\ion{Al}{3}$\lambda1990$ flux.}
\tablenotetext{i}{Based on the range of values of H$\alpha/$H$\beta$ (3.97--6.64) 
	from the NLS1 sample of \citet{OP85}.}
\tablerefs{(1) This work. (2) Paper~V. (3) Paper~IV. (4) Paper~II. (5) Paper~III.}
\end{deluxetable}
\clearpage 

\begin{deluxetable}{llcccc}	
\tablewidth{0pc}
\tablecaption{Spectral Indices\label{sa:aox}}    
\tablehead{\colhead{Index}  & \colhead{Definition} & 
	   \colhead{Ark~564\tablenotemark{a}} & \colhead{Ton~S180\tablenotemark{a}} 
				& \colhead{BLSy1} & \colhead{References} \\ 
	   \colhead{(1)} &\colhead{(2)} & \colhead{(3)} & \colhead{(4)} & \colhead{(5)}  & \colhead{(6)}
}
\startdata   
$\alpha_{\rm 100\,\mu m-12\,\mu m}$  & -1.086 log($F_{\rm 12\,\mu m}/F_{\rm 100\,\mu m})$ & 0.64& \\
$\alpha_{\rm  12\,\mu m-2.4\,\mu m}$  & -1.431 log($F_{\rm 2.4\,\mu m}/F_{\rm 12\,\mu m})$ & 1.64& \\
$\alpha_{\rm 2.4\,\mu m-1.6\,\mu m}$  & -5.679 log($F_{\rm 1.6\,\mu m}/F_{\rm 2.4\,\mu m})$  & 1.94& \\
$\alpha_{\rm 1.6\,\mu m-1\,\mu m}$  & -4.900 log($F_{\rm 1\,\mu m}/F_{\rm 1.6\,\mu m})$  & 1.18\tablenotemark{b}& \\
$\alpha_{\rm 12\,\mu m-1\,\mu m}$  & -0.927 log($F_{\rm 1\,\mu m}/F_{\rm 12\,\mu m})$  & 1.60\tablenotemark{b} &  &  & \\
$\alpha_{\rm 3000\,\AA-1000\,\AA}$ ($\alpha_{\rm uv}$) & -2.096 log($F_{\rm 1000\,\AA}/F_{\rm 3000\,\AA})$ 
	& 0.42 & 0.66 & 1.25 & 1 \\ 
$\alpha_{FUSE-ASCA}$\tablenotemark{c}  & & 1.08 &  &  & 1 \\ 
$\alpha_{\rm 5500\,\AA-0.25\,keV}$  	              & -0.489 log($F_{\rm 0.25\,keV}/F_{\rm 5500\,\AA})$ 
        & 0.82\tablenotemark{b} & 1.12 & 0.73 & 2 \\
$\alpha_{\rm 5500\,\AA-1\,keV}$ ($\alpha_{\rm ox-hard}$) & -0.378 log($F_{\rm 1\,keV}/F_{\rm 5500\,\AA})$ 
	& 0.93 & 1.38 & 1.13 & 3 \\
$\alpha_{\rm 1\mu m -2\,keV}$ ($\alpha_{\rm ix})$   & -0.312 log($F_{\rm 2\,keV}/F_{\rm 1\mu})$ 
	& 1.01\tablenotemark{b} & 1.35 & 1.14-2.16 & 4 \\
$\alpha_{\rm 2500\,\AA-2\,keV}$ ($\alpha_{\rm ox})$ & -0.384 log($F_{\rm 2\,keV}/F_{\rm 2500\,\AA})$  
	& 1.11 & 1.52 & 1.46$^{+0.05}_{-0.07}$, 1.21$\pm$0.02 & 5,6 \\
$\alpha_{\rm x}$                            & -1.431 log($F_{\rm 10\,keV}/F_{\rm 2\,keV})$  
	& 1.57 & 1.44 & 0.91 & 7
\enddata
\tablecomments{For spectral indices relative to the Ark~564 {\it IRAS} points
	we used the reddening-corrected, redshift-corrected fluxes 
	and wavelengths.}
\tablenotetext{a}{Intrinsic, i.e, reddening-corrected, redshift-corrected (SEDA).}
\tablenotetext{b}{Based on extrapolated value (\S\ref{sa:sed}, Figure~\ref{sa:sedcorr}).}
\tablenotetext{c}{Simple power law connecting the high energy end of the {\it FUSE} 
	spectrum and the low energy end of the {\it ASCA} spectrum 
	(\S\ref{sa:sed}, Figure~\ref{sa:sedcorr}).}
\tablerefs{
(1) \citealt{CGK91}; index in the 2200-1200 \AA\ band, 
based on the BLS1 sub-sample. (2) \citealt{tea99c}. (3) \citealt{gp98}.
(4) \citealt{law97}. (5) \citealt{zam81}. (6) \citealt{puch96}. (7) \citealt{Nandraea97b}.  
}
\end{deluxetable}

\begin{deluxetable}{lcccc}	
\tablewidth{0pc}
\tablecaption{Luminosities\label{sa:enbudget}}    
\tablehead{\colhead{Energy Range}  & \colhead{$L$ (Observed)}  & \colhead{$L$ (SEDA)\tablenotemark{a}} 
		         & \colhead{$L$ (SEDB)\tablenotemark{a}} 
		         & \colhead{$L$ (XMM)\tablenotemark{a}}    \\
	   \colhead{(keV)} & \colhead{($\times 10^{44}$\,\ergsec)} & \colhead{($\times 10^{44}$\,\ergsec)} 
		        & \colhead{($\times 10^{44}$\,\ergsec)} & \colhead{($\times 10^{44}$\,\ergsec)} \\ 
	   \colhead{(1)} & \colhead{(2)} & \colhead{(3)}  & \colhead{(4)}  & \colhead{(5)}  
}
\startdata   
$10^{-5}$--$ 10^{-4}$	&\nodata 	&  2.1192 	&  2.1192	&  2.1192  \\
$10^{-4}$--$ 10^{-3}$	&\nodata 	&  1.8108	&  1.8108	&  1.8108\\
$10^{-3}$--0.01	&\nodata 	&  1.2464	&  1.2464	&  1.2464\\
0.01--0.1		& \nodata 	&  3.5175	& 2.2589	& 4.1430 \\
0.1--1     	      	& 0.2201  	&  1.6314	& 1.8080	&2.7716 \\ 
1--10       	    	& 0.4086	&  0.4485	& 0.4485	& 0.5285\\
$10^{-5}$--10		& \nodata	& 10.7738	& 9.6918	&12.6202 
\enddata
\tablenotetext{a}{Reddening-corrected, rest-frame luminosities.}
\tablecomments{The SEDA luminosities have been calculated using 
power-law parameterization of the SED with the spectral indices 
reported in Table~\ref{sa:aox}. 
The SEDB luminosities refer to the more conservative 
parameterization described in \S\ref{sa:sed} (Figure~\ref{sa:sedcorr}). 
The XMM luminosities make use of the {\it XMM} spectrum in the 0.05--10\,keV
band.
    }
\end{deluxetable}

\end{document}